# Transport measurements in twisted bilayer graphene: Electron-phonon coupling and Landau level crossing

Ting-Fung Chung,[1,*] Yang Xu,[1,*] and Yong P. Chen[1,2,3,†]

[1]*Department of Physics and Astronomy and Birck Nanotechnology Center, Purdue University, West Lafayette, Indiana 47907, USA*

[2]*School of Electrical and Computer Engineering and Purdue Quantum Center, Purdue University, West Lafayette, Indiana 47907, USA*

[3]*WPI-AIMR International Research Center on Materials Science, Tohoku University, Sendai 980-8577, Japan*

[*]These authors contributed equally to this work.

[†]yongchen@purdue.edu

## Abstract

We investigate electronic transport in twisted bilayer graphene (tBLG) under variable temperatures (*T*), carrier densities (*n*), and transverse magnetic fields, focusing on samples with small-twist-angles ($\theta$). These samples show prominent signatures associated with the van Hove singularities (VHSs) and superlattice-induced mini-gaps (SMGs). Temperature-dependent field effect measurement shows that the difference between temperature-dependent resistivity and residual resistivity, $\rho_{xx}(T,n)-\rho_0(n)$, follows $\sim T^{\beta}$ for *n* between the main Dirac point (DP) and SMG. The evolution of the temperature exponent $\beta$ with *n* exhibits a W-shaped dependence, with minima of $\beta$~0.9 near the VHSs and maxima of $\beta$~1.7 toward the SMGs. This W-shaped behavior can be qualitatively understood with a theoretical picture that considers both the Fermi surface smearing near the VHSs and flexural-acoustic phonon scattering. In the quantum Hall regime, we observe only Landau level crossings in the massless Dirac spectrum originating from the main DP but not in the parabolic band near the SMG. Such crossings enable the measurement of an enhanced interlayer dielectric constant, attributed to a reduced Fermi velocity. Moreover, we measure the Fermi velocity, interlayer coupling strength, VHS energy relative to the DP, and gap size of SMG, four important parameters used to describe the peculiar band structure of the small-$\theta$ tBLG.

**Introduction**

Twisted bilayer graphene (tBLG), which can be formed by stacking two graphene crystals with a twist angle ($\theta$), is an important example of moiré crystals[1–8]. The tBLG with small-$\theta$ is particularly interesting, since the moiré pattern periodicity enlarges and the separation between the van Hove singularity (VHS) and Dirac point (DP) shrinks when reducing $\theta$, yielding dramatic changes to the electronic band structure near the DP. In earlier transport studies[9–12], however, sample disorder and limited tunability in the carrier density (e.g., by ~$6 \times 10^{12}$ $cm^{-2}$ for typical $SiO_2$/Si backgates) hindered the investigation of the electrical properties of small-$\theta$ tBLG. Recent advances in the accurate manipulation of $\theta$ (down to $\leq 2°$) and high-quality tBLG samples sandwiched between two layers of hexagonal boron nitride (h-BN) have revealed many intriguing transport features associated with tBLG and its moiré band[3], such as VHSs[13–17], superlattice-induced mini-gaps (SMGs)[16,17], magnetic-field-induced Hofstadter butterfly spectrum[18], and Fabry-Pérot interferences due to networks of helical states formed between the alternating AB/BA regions in very small-$\theta$ tBLG[19,20]. Particularly, recent experiments performed on tBLG near the “magic-angle” (~1.1°)[21,22] revealed that tBLG can exhibit flat energy band near charge neutrality and Mott-like insulating states at half-filling[21] as well as superconducting domes when the carrier density is slightly away from the half-filled case[22].

Although prominent transport signatures[15–18] related to the VHS and SMG have been reported for h-BN-sandwiched tBLG samples with $\theta \leq 2°$, there remain many open questions regarding the transport characteristics of this system. One is that little is known about the electron-phonon (el-ph) coupling as a function of temperature (*T*) and carrier density (*n*) for in-plane transport and in particular, how the VHS and SMG alter the el-ph coupling. Acoustic phonon-contributed resistivity and phonon-limited carrier mobility have been extensively studied in monolayer and Bernal (AB)-stacked bilayer graphene[23–26]. However, for tBLG, thus far such experiments have been performed only for interlayer transport and the samples with relatively large-$\theta$ [27,28]. Another is regarding the measured (transport) gap of SMG. Earlier tBLG devices fabricated on $SiO_2$/Si did not show a mini-gap[9,11,12], while in the h-BN-encapsulated samples, it has been observed that the gap size of SMG varies widely from ~10–60 meV for $\theta \sim 1.8°–2°$.[16,17] Further, the nature of this mini-gap, which is found to be several times larger than the theoretical prediction[17], remains to be fully understood.

Here, we report on a transport study of top- and back-gated tBLG samples with h-BN encapsulation under variable *T*, *n* and magnetic fields (*B*). In this study, we focus on tBLG with small-$\theta \sim 2°$ (but still larger than the magic angle)[21]. Our high-quality tBLG devices, exhibiting notable transport features corresponding to the VHSs and SMGs, confirm the recent finding of relatively large SMG gap, and provide new insights into the acoustic phonon scattering and interlayer coupling in the small-$\theta$ regime. We observe the *T*-dependence of acoustic phonon-contributed resistivity at various *n* follows a power-law, $\sim T^{\beta}$. The *T*-exponent $\beta$ of the resistivity shows a W-shaped *n*-dependence and evolves from ~0.9 to ~1.7 when tuning *n* away from the VHS. Additionally, as we adjust the transverse electric field (interlayer potential) in the samples in the quantum Hall (QH) regime, a mapping of the Landau quantization shows crossings of two sets of Landau levels (LLs) for *n* below the VHS but only one set of LLs (no crossing) for *n* beyond the VHS. By analyzing the electric field-induced LL crossings, we find enhanced interlayer screening in tBLG (the interlayer dielectric constant is ~6 times of the vacuum permittivity), which is understood as a consequence of the reduced Fermi velocity ($v_F$) due to the interlayer interaction. We also deduce the interlayer coupling strength, VHS energy ($E_{VHS}$, the energy difference between the main DP and VHS), SMG gap size, and the reduced $v_F$, revealing strong interlayer coupling in our h-BN-sandwiched tBLG and providing an interpretation consistent with recent scanning tunneling microscopy (STM) findings[15,29] and calculations[15,30].

**Results and discussion**

Our samples consist of h-BN/tBLG/h-BN stacks, focusing on small-$\theta$ around 1.3°–2°, and an intermediate $\theta$ ~5° as a reference. We assembled tBLG using the dry transfer method[31,32]. The angle alignment was achieved by breaking and stacking from the same large piece of single crystal graphene flake (exfoliated from Kish Graphite from Covalent Materials Corp.) on a rotary stage with angular accuracy ~0.1°, as depicted in Fig. 1(a-d). Figure 1(g) shows representative Raman spectra (measured with a 638 nm laser excitation) of three samples ($\theta$ ~1.4°, 2° and 5°) before thermal annealing (post-annealed samples used in our devices show noisier, but qualitatively similar Raman spectra, see Fig. S1 for details). We observe a broadening of the G band and an asymmetric 2D band when reducing $\theta$, similar to the prior report in double-layer graphene (using stacks of chemical vapor deposition (CVD)-grown graphene)[33]. These $\theta$-dependent Raman features indicates the tBLG samples with a relatively small-$\theta$ (accurate determination of $\theta$ is by transport measurement as described in Fig. 2)[33]. The stack is patterned into an edge-contacted device[31]. The device has both top and back gates for controlling the total carrier density $n = n_L + n_U$ (where $n_L$ and $n_U$ is the carrier density of the lower and upper graphene layers, respectively) and the average displacement field (applied normal to the layers) $D$ between the two layers. By adjusting both gates, we can separately tune $n=\left(C_{\mathrm{B}}\Delta V_{\mathrm{BG}}+C_{\mathrm{T}}\Delta V_{\mathrm{TG}}\right)/e$ and $D=\left(C_{\mathrm{B}}\Delta V_{\mathrm{BG}}-C_{\mathrm{T}}\Delta V_{\mathrm{TG}}\right)/2$, where $C_{\mathrm{T(B)}}$ is the capacitance per unit area of the top- (back-) gate dielectric, $e$ = 1.602 × $10^{-19}$ C is the elementary charge, $\Delta V_{\mathrm{T(B)G}}=V_{\mathrm{T(B)G}}-V_{\mathrm{T(B)G}}^{\circ}$, $V_{\mathrm{T(B)G}}$ is the applied top- (back-) gate voltage, and ($V_{\mathrm{TG}}^{\circ}$, $V_{\mathrm{BG}}^{\circ}$) are the gate voltages when both upper and lower graphene layers are charge neutral, and $D=0$ indicates $n_{\mathrm{L}}=C_{\mathrm{B}}\Delta V_{\mathrm{BG}}/e=C_{\mathrm{T}}\Delta V_{\mathrm{TG}}/e=n_{\mathrm{U}}$. The simple approximations for $n$ and $D$ above are good because the quantum capacitance of doped graphene is at least an order of magnitude larger than the gate capacitance (with the h-BN layer as a gate dielectric) thus can be neglected. In Device A (see Fig. 2(a)), for example, we obtain $V_{\mathrm{TG}}^{\circ}$ = -1.45 V and $V_{\mathrm{BG}}^{\circ}$ = 32.8 V (corresponding to the intersection of the two dashed arrows indicating axes of $n$ and $D$). The gate capacitances are calculated from the thicknesses of h-BN and $SiO_2$ and are confirmed with gate-dependent Hall measurements. Details of sample preparation and device fabrication are in Supplemental Material[34].

We focus on Device A ($\theta$~2°), which shows a Hall mobility ~25,000 $cm^2V^{-1}s^{-1}$ for $n \approx 1.5 \times 10^{12}$ $cm^{-2}$ at $T$ = 1.6 K. Data from other tBLG devices are presented in Fig. S2. A measurement of the (four-terminal) longitudinal resistance ($R_{xx}$) versus $V_{TG}$ and $V_{BG}$ is shown in the color plot in Fig. 2(a), taken at $B$ = 0 T and $T$ = 1.6 K. The central blue stripe denotes the resistance peak of total CNP in which both layers have equal and opposite carrier densities such that the tBLG maintains charge neutral (total $n$ = 0). The resistance of the central CNP as a function of $D/\varepsilon_0$ (with $\varepsilon_0$ being the vacuum permittivity) is displayed in the inset, showing that the resistance is reduced by a factor of ~2 as $D$ increases, similar to that in large-$\theta$ tBLG (see Fig. S2(c) and Ref. [35]). In addition to the central CNP, two parallel red stripes (which are relatively insulating) away from the CNP are e-SMG and h-SMG (here e- and h- denote electron- and hole-side, respectively)[16,17]. The resistance of the SMG is tunable by $D/\varepsilon_0$, as depicted in the h-SMG with reducing resistance (color from red to yellow) at larger *D*. This reduction in the resistance of SMG with *D* could be understood as a result of lifting the subband degeneracy due to the interlayer potential[36].

Figure 2(b) presents $R_{xx}$ and the Hall resistance ($R_{xy}$) measured as functions of *n* along the dashed line in (a) at *B* = 1 T and *T* = 1.6 K. We observe three abrupt zero crossings in $R_{xy}$, where $R_{xx}$ also reaches maximum, at $n = 0$ and $n = n_s \approx \pm 9.9 \times 10^{12}$ cm$^{-2}$ (corresponding to CNP and SMGs, respectively, represented by the blue stripe and two red stripes in (a)). The gradual sign reversal in $R_{xy}$ at $n = n_{VHS} \approx \pm 5 \times 10^{12}$ cm$^{-2}$ accompanied by a shallow resistance peak in $R_{xx}$ are attributed to the VHS. From the carrier density ($n_s$) at SMG[8], we can estimate the superlattice unit cell area $A_{ms} = 4/n_s \approx 40.4$ nm$^2$ and the superlattice wavelength $\lambda_{ms} = \left(2A_{ms}/\sqrt{3}\right)^{1/2} \approx$ 6.8 nm. According to $\lambda_{ms} = \frac{a}{2\sin(\theta/2)}$, where $a$ = 0.246 nm is the lattice constant of graphene, we obtain $\theta \sim 2^\circ$ (consistent with the intended value in the fabrication and the estimate based on the Raman measurement). Our device has a notably different resistance of SMG compared to the devices of similar $\theta$ (~2°) in recent studies[16,17].

Figure 2(c) displays *T*-dependence of $R_{xx}$ (at *B* = 0 T) for Device A measured along the dashed line in (a). Note that a small variation of *D* is present along the dashed line due to limitations in the gate voltage to access both SMGs. The resistance of both SMGs increases by about an order of magnitude, accompanied by a narrowing of the resistance peak, as *T* decreases from 300 K to 40 K. We extract the resistance of both SMGs at $\pm n_s$ for various *T* and plot the log of conductance ($G_{xx} = 1/R_{xx}$) versus 1/*T*, as shown in Fig. 2(d). The h-SMG's $G_{xx}$ (open squares) decreases slightly faster than that for the e-SMG (open circles), but both appear to begin saturating below ~30 K. It is evident that the SMG's $G_{xx}$ above 120 K follows the thermally-activated behavior, $G_{xx} \propto \exp(-\Delta/2k_BT)$, where $\Delta$ is the thermal activation (TA) gap, $k_B$ is the Boltzmann constant. At lower $T$, the deviation from the thermally-activated transport to the much weaker $T$-dependence is attributed to the Mott variable range hopping (VRH) conduction mediated by localized states. These localized states are attributed to disorder, as indicated by the limited Hall mobility, and to adjacent high energy bands accessible by phonon-assisted indirect transitions[16,27,37]. We thus add an extra term to represent the Mott VRH conductance and fit our data (over the temperature range between 15 K and 300 K) to $G_{xx} = G_{TA}e^{-\Delta/2k_BT} + G_{VRH}e^{-(T_o/T)^{1/3}}$, where $G_{TA}$ and $G_{VRH}$ are the prefactors of TA and VRH terms, respectively, and $T_o$ is the characteristic temperature for VRH. For the e- and h-SMGs, we find $\Delta$ ~65 meV and ~45 meV, respectively. We measured $\Delta$ (~52 – 79 meV) in two more devices with $\theta<2°$ (see Fig. S3 for the fits and $\Delta$ for all three devices with $\theta$~1.3°–2°).

Recent reports on small-$\theta$ tBLG have found a range of $\Delta$ for the superlattice-induced insulating behavior. Our experimentally measured $\Delta$ are comparable to the results (50–60 meV) reported in Ref. [17], which are 5–10 times higher than those in earlier experiments and theoretical calculations[16,17]. Several reasons have been proposed to explain this surprisingly large $\Delta$ measured in experiments (nearly ~10 times larger than the calculated $\Delta$), such as the formation of domains of different stacking and lattice deformation (strain), buckling effect, many-body interactions, and under-estimated interlayer coupling strength ($t_\theta$)[17,38,39]. We rule out the unexpectedly large $t_\theta$ from our analysis of magnetotransport measurements discussed below. The obtained $t_\theta$ is found to be comparable to previous calculations and STM results[1,12,29,40]. Precise causes for the large $\Delta$ remain to be better understood.

Figure 2(e) shows $T$-dependence of the longitudinal resistivity ($\rho_{xx}$, sheet resistivity) for several $n$ between the CNP and e-SMG, corresponding to the range marked by the dashed rectangle in Fig. 2(c). We find that for each measured $n$ between $2 \times 10^{12}$ and $8 \times 10^{12}$ cm$^{-2}$, $\rho_{xx}(T)$ decreases with decreasing $T$ (metallic behavior, $d\rho_{xx}/dT > 0$, attributed to acoustic phonon scattering) and saturates (below 20 K) to a residual value $\rho_0(n)$ ~$(115 \pm 35)\ \Omega$ (or $(4.5 \pm 1.3) \times 10^{-3}$ h/e$^2$), attributed to charged impurity scattering. The observed metallic behavior is $n$-dependent, showing a different rate of resistivity increase with increasing $T$. Similar results of the sample for $n < 0$ (between the CNP and h-SMG) are presented in Fig. S4(b). In contrast to the tBLG, monolayer graphene exhibits a linear temperature dependence in resistivity ( $\rho_{xx} \propto T$ ), independent of $n$, and AB-bilayer graphene shows very weak $T$-dependence over comparable $n$ ranges as we measured[25,41]. We have also examined $T$-dependent $\rho_{xx}$ of the reference Device D ($\theta$ ~5°, see Fig. S5). The Dirac cones of those bilayers are displaced by a large wavevector in momentum space and mostly decoupled. Hence, the VHSs ($\pm n_{VHS}$) of such samples are out of the range of accessible $n$. In Device A, we find that the room temperature resistivity is higher than the low-$T$ saturation value by $\rho_{xx}(n,T\text{=}300\ \text{K}) - \rho_0(n)$ ~300–500 $\Omega/\square$, attributed to the contribution due to electron-acoustic phonon scattering. In contrast, $\rho_{xx}(n,T\text{=}300\ \text{K}) - \rho_0(n)$ is only ~30 $\Omega/\square$ in Device D (Fig. S5) over comparable ranges of $n$. This difference may be attributed to that Device D has a larger separation of the Dirac cones from the upper and lower graphene layers in momentum space, $\Delta K = 2\left|\Gamma \mathrm{K}\right|\sin\left(\theta/2\right)$, where $\Gamma \mathrm{K} =$ 1.703 Å$^{-1}$ being the distance between the $\Gamma$ and K points of graphene Brillouin zone, thus requiring phonons with larger momentum (compared to Device A) to couple electrons between the layers.

To quantitatively discern the difference in the resistivity of the tBLG at various $n$, we fit the $\rho_{xx}(T<\sim 150\text{ K})$ data to $\Delta\rho_{xx} = \rho_{xx}(n,T) - \rho_0(n) = \alpha T^{\beta}$, where $\alpha$ is the prefactor and $\beta$ is the $T$-exponent. Figure 2(f) presents $\beta$ versus $n$ for Devices A ($\theta\sim 2°$) and D ($\theta\sim 5°$). The $\beta$ value of Device A displays a W-shaped curve with minima of ~0.9 at $\pm n_{VHS}$ and maxima of ~1.4–1.6 when $n$ approaches to $\pm n_{SMG}$, whereas for Device D, $\beta$ ranges ~1–1.3 and does not show a strong dependence on $n$. Note that the measured $\beta$ differs from that in monolayer graphene in which the in-plane acoustic (LA/TA) phonon scattering gives rise to a linear-in-$T$ resistivity ($\beta \approx 1$)[23,25]. The resistivity of tBLG, however, can be significantly affected by both interlayer scattering via flexural phonons and intralayer scattering via in-plane acoustic phonons[27,37,42,43], leading to $\beta > 1$, as observed in Device A and Device D (in the regime of two decoupled monolayers). While one might expect similar phonon scattering scenario for both devices, the characteristic band structure of tBLG in the regime of small-$\theta$ as in Device A could markedly affect the resistivity. Near the VHSs, a suppression of $v_F$ caused by the interlayer coupling[43] leads to a rise in the resistivity, manifested as the small and broad peaks located at $\sim\pm n_{VHS}$, as shown in Fig. 2(c). At higher $T$, thermal broadening[43] smears out these resistivity peaks and decreases $\beta$ to ~0.9. Theories[37,43] have considered different contributions of acoustic phonon modes to the el-ph scattering in tBLG at various $\theta$. The theories[37,43] have predicted a significant change in the contribution of different phonon modes to the resistivity when $n$ increases toward SMG in the small-$\theta$ regime, which may offer an interpretation for the distinct $n$-dependence of $\beta$ (Fig. 2(f)) we observed in Devices A and D.

We have also measured quantum Hall (QH) effects (QHE) in such small-$\theta$ tBLG samples as Device A and found features different from those in either AB-bilayer or large-$\theta$ tBLG[35,44]. Figure 3(a) presents a color plot of $R_{xx}$ versus $V_{BG}$ and $V_{TG}$ for Device A, acquired at $B$ = 6 T and $T$ = 1.6 K. The central and side white stripes represent the CNP and SMGs, located at the same positions as those shown in Fig. 2(a) measured at $B$ = 0 T. In the plot, we can observe two markedly different types of LL-like structures originating from the CNP (total $n$ = 0) and the side SMGs, separated by the VHSs (white dashed lines). As we will discuss in the following, the LL crossings observed in the vicinity of CNP (between e-VHS and h-VHS) stem from two sets of LLs of the graphene bilayers when $D$ lifts the layer degeneracy, similar to that observed in large-$\theta$ tBLG[35]. On the other hand, we observe only one set of LLs that manifests as lines parallel to those corresponding to CNP and SMG (dashed lines in Fig. 3(a)), for $n$ beyond e- or h-VHS.

The zoomed-in resistance map $R_{xx}$ ($B$ = 6 T) from the region enclosed by the blue solid lines in (a) is shown in Fig. 3(b). The gate voltages are converted to $D/\varepsilon_0$ and filling factor (measured from the e-SMG) $\nu_{\mathrm{e}} = \left(n - n_{\text{e-SMG}}\right) h/eB$, where $h$ is the Planck's constant. The negative values of $\nu_{\mathrm{e}}$ in Fig. 3(b) denote hole-like carriers between e-SMG and e-VHS (also see $R_{xy}$ in Fig. 2(b)). We observe the sequence of the QH states (black stripes) following steps of 4 in $\nu_{\mathrm{e}}$ (i.e., –4, –8, –12, ...), which is independent of $D$. Figure 3(c) displays the $R_{xx}$ and $R_{xy}$ as functions of $n$ at $B$ = 6 T, measured along the orange dashed line with $D/\varepsilon_0$ = -0.51 V/nm in (b). $R_{xy}$ exhibits several developing quantized plateaus at $-h/8e^2$, $-h/12e^2$, $-h/16e^2$, accompanied by minima in $R_{xx}$. The $\nu_{\mathrm{e}}$ sequence indicates massive fermions (attributed to the parabolic bands around the e-SMG at the $\Gamma_s$ point of the superlattice Brillouin zone[16,17]) and the 4-fold degenerate LLs, which follow from the spin degeneracy and "Fermi contour degeneracy" of the parabolic energy band near the e-SMG[16,17]. We also measure Shubnikov-de Haas (SdH) oscillations at fixed gate voltages (denoted by the green open square in both (a) and (b)), as shown in the inset of (b). The Landau plot (LL index ($N$) *vs* $1/B$) of the oscillations in the inset reveals a zero $N$-intercept. This is an indication of zero Berry phase, which is another key feature that is different from the massless charge carriers in monolayer graphene.

We now turn to the CNP region ($n$ ~0), showing LL crossings emanated from the lower and upper graphene layers. Figure 4(a) displays the zoomed-in color plot of $R_{xx}$ ($B$ = 6 T) between the two VHSs (white dashed lines) in Fig. 3(a) as a function of $D/\varepsilon_0$ and $\nu = (n/B)h/e$. The filling factor combination $\nu = \nu_L + \nu_U$ for several expected QH states (regions in black) has been labeled as a guide to the eye. The subscripts L and U represent lower and upper graphene layers, respectively. A complete set of $\nu$ for all expected QH states (according to Ref. [35]) is schematically illustrated in Fig. 4(b). Figure 4(c) presents the $R_{xx}$ and $R_{xy}$ versus $n$ measured at $D$ = 0 (along the orange dashed line in (a)). We observe developing quantized plateaus in $R_{xy}$ at $\pm h/\nu e^2$ for $\nu$ = 4, 12, 20, 28 (with steps of 8), consistent with the $\nu$ assignment for both electron (+ sign) and hole (– sign) doping in (b). This 8-fold degeneracy arises from the spin, valley and layer degeneracies of two monolayers[35]. A similar set of developing plateaus in $R_{xy}$ is observed in the magnetic field dependent data at $D$ = 0 and $n = 2.4 \times 10^{12}$ cm$^{-2}$, as in Fig. 4(d). We assign each minimum in $R_{xx}$ of the oscillations to its corresponding $\nu = 8(N + 1/2)$, where $N$ = 0, ±1, ±2, … is the index of the filled LL in each graphene layer (noting the lower and upper graphene layers are degenerate with the same filling at $D$ = 0). The data of $N$ *vs* 1/$B$ can be linearly fitted with slope 12.7 T and intercept ~-0.5 in the vertical $N$-axis (see the inset in Fig. 4(d)), revealing a Berry phase $\pi$ attributed to the decoupled monolayer graphene each possessing a carrier density of $n$/2. On the other hand, we see an alternating stripe pattern (i.e., LL crossing) with changing $D$ in Fig. 4(a), as expected from two decoupled monolayers[10,35]. We further observe a beating pattern in the SdH oscillations at $D/\varepsilon_0$ = -1.2 V/nm (see Fig. 4(e)), confirming a superposition of two independent sets of QH states with different filling factors from the two decoupled monolayers. The inset presents the Fourier transform (FT) amplitude versus frequency corresponding to the data (when plotted as $R_{xx}$ *vs* 1/B) in Fig. 4(e), exhibiting two prominent peaks arising from the carrier densities of different layers ($n_U$ ~$2.9 \times 10^{12}$ cm$^{-2}$ and $n_L$

~8.1 × 10[11] cm[-2]). These results indicate that the low-energy electronic structure ($n < n_{\mathrm{VHS}}$) of tBLG ($\theta$~2°) corresponds to that of two decoupled graphene monolayers.

Next, we have performed *T*-dependent SdH oscillation studies in the decoupled regime in Device A. We estimate the cyclotron mass ($m^*$) as well as $v_{\mathrm{F}}$ from the *T*-dependent oscillations at $n = 1.4 \times 10^{12}$ cm$^{-2}$ and $D = 0$, where the DP of two layers (with comparable doping) is vertically aligned and the band renormalization caused by the interlayer asymmetric potential is negligible[12,45]. Figure 4(f) shows the *T*-dependence of the oscillation amplitude $\Delta R_{xx}$ at $n = 1.4 \times 10^{12}$ cm$^{-2}$. The $\Delta R_{xx}$ for the oscillation at 0.2 T$^{-1}$ ($\nu = 6+6$ QH state) is normalized by the $\Delta R_{xx}(T = 1.6$ K) and is displayed in the inset as a function of *T*. By fitting to the Lifshitz-Kosevich formula[46], $\Delta R_{xx}(T,B) \propto \left[\left(\chi T/\hbar\omega_c\right)/\sinh\left(\chi T/\hbar\omega_c\right)\right]e^{-\chi T_D/\hbar\omega_c}$, where $\chi$ is a constant, $\hbar\omega_c = \hbar eB/m^*$ and $T_D$ are the fitting parameters, we can extract $m^*$ ~0.029$m_e$ (with $m_e$ being the electron rest mass) at the Fermi energy. With the Onsager relation, $B_F = A_k\left(\Phi_o/2\pi^2\right)$, we can extract the Fermi momentum ($k_{\mathrm{F}} = \sqrt{A_F/\pi}$; a circular Fermi surface $A_k$ of Dirac cone is assumed when the Fermi energy is close to the main CNP and away from VHS) from the SdH oscillation frequency ($B_F$ ~6.85 T) obtained in Fig. 4(f) and then $v_{\mathrm{F}} = \hbar k_{\mathrm{F}}/m^*$ ~0.58 × 10$^6$ ms$^{-1}$, about a 40% reduction compared with that in monolayer graphene ($v_{\mathrm{F}}^0 \approx 10^6$ ms$^{-1}$). The reduced $v_{\mathrm{F}}$ is consistent with the finite interlayer coupling in the small-$\theta$ tBLG, possessing both low-energy VHSs and SMGs. We also measure a similar $v_{\mathrm{F}}$ ~(0.56 ± 0.02) × 10$^6$ ms$^{-1}$ at $n = 2.5 \times 10^{12}$ cm$^{-2}$ (see Fig. S6). In addition to Device A, we performed similar measurement on Device D (with $\theta$ ~5°) at similar carrier densities and obtained $v_{\mathrm{F}}$ ~1 × 10$^6$ ms$^{-1}$ (see Fig. S7), comparable to the monolayer graphene value $v_{\mathrm{F}}^0$. Our results confirm that $v_{\mathrm{F}}$ depends strongly on both $\theta$ and interlayer coupling in tBLG.

From the reduced $v_{\mathrm{F}}$, we can estimate the $t_{\theta}$ and VHS energy ($E_{\mathrm{VHS}}$, the energy difference between the main DP and VHS) — parameters reflecting the interlayer interactions in our tBLG encapsulated in h-BN. It has been shown that $v_{\mathrm{F}}$ decreases with decreasing $\theta$ or increasing interlayer coupling strength ($t_{\theta}$)[40], $v_{\mathrm{F}}/v_{\mathrm{F}}^{0}=1-9\left(t_{\theta}/\hbar v_{\mathrm{F}}^{0}\Delta K\right)^{2}$, where $\Delta K$ is the separation between the two DPs ($\mathrm{K_L}$ and $\mathrm{K_U}$) in momentum space and ħ is the reduced Planck constant. For $\theta$ = 2°, $\Delta K$ = 0.059 Å$^{-1}$, and $v_{\mathrm{F}}$ = 0.58 $v_{\mathrm{F}}^{0}$, we obtain $t_{\theta}$ = (84 ± 5) meV, which is in a good agreement with prior theoretical and experimental studies[1,12,15,29,40]. We note the similarity of $t_{\theta}$ measured from small-$\theta$ tBLG on different substrates ($SiO_2$[12] and h-BN (this work)), suggesting that $t_{\theta}$ is relatively insensitive to the surrounding dielectric environment, interfacial strain, and disorder. The energy difference between the two VHSs can be estimated[1] by:

$$\Delta E_{\mathrm{VHS}}=2E_{\mathrm{VHS}}\approx\hbar v_{\mathrm{F}}^{0}\Delta K-2t_{\theta}\,. \qquad (1)$$

By assuming $t_{\theta}$ is comparable in the e-doped and h-doped sides, the equation above yields

$\Delta E_{\mathrm{VHS}}=2E_{\mathrm{VHS}}$ ~220 meV and $E_{\mathrm{VHS}}$ ~110 meV.

In our experiment, we can also deduce the $E_{\mathrm{VHS}}$ from the Landau quantization pattern ($R_{xx}$ *vs* $V_{TG}$ and $V_{BG}$ at $B$ = 6 T), as presented in Fig. 3(a). Below the VHSs, the tBLG behaves like two decoupled graphene layers. The LL energy[46] of each monolayer graphene with (reduced) $v_{\mathrm{F}}$ in perpendicular magnetic field $B$ is given by:

$$E_{\mathrm{N}} = \mathrm{sgn}(N)\sqrt{2e\hbar v_{\mathrm{F}}^{2}B|N|}\ , \qquad (2)$$

where *N* is the corresponding LL index and the $R_{xx}$ minima would occur at $\nu = 4(N + 1/2)$. As presented in Fig. 3(a) (also see Fig. S8(a)), equally-spaced lines (passing through points of equal filling factors $\nu = nh/eB = 2$, 6, 10 … in the two layers) parallel to those corresponding to CNP and VHS can be defined. We find e-VHS is located near *N* ~3.5 ($\nu$ ~16 for monolayer), which yields $E_{\mathrm{VHS}}$ ~(95 ± 4) meV (calculated from the LL energy expression with a value $v_{\mathrm{F}}$ ~0.58 $v_{\mathrm{F}}^{0}$). This is in a reasonable agreement with $E_{\mathrm{VHS}}$ ~110 meV extracted from Eq. (1) above. Our extracted $E_{\mathrm{VHS}}$ values are also consistent with a recent STM study of CVD tBLG on h-BN substrate[15].

We further investigate the effect of the reduced $v_{\mathrm{F}}$ on the interlayer screening of the tBLG. Close to the DP, the density of states vanishes, causing the tBLG to become less efficient in screening adjacent electric fields[47]. The incomplete charge screening creates a charge density imbalance (Δ*n*) as well as an interlayer potential difference (Δ*V*) between the two graphene layers. The interlayer potential difference with an interlayer spacing ($d_{\mathrm{GG}}$) depends on the difference between the average displacement field (*D*) and the screening field (eΔ*n*/2),[35]

$$-\Delta V = (D - e\Delta n/2)/C_{\mathrm{GG}}\ , \qquad (3)$$

where $C_{\mathrm{GG}} = \varepsilon_0 \varepsilon_{\mathrm{GG}}/d_{\mathrm{GG}}$ is the interlayer capacitance per unit area and $\varepsilon_{\mathrm{GG}}$ is the interlayer dielectric constant. When two LLs (one from the lower layer with index $N_{\mathrm{L}}$, the other from the upper layer with index $N_{\mathrm{U}}$) cross, the LL energy difference ($E_{N_{\mathrm{L}}} - E_{N_{\mathrm{U}}}$) between them provides a measure of Δ*V*, $E_{N_{\mathrm{L}}} - E_{N_{\mathrm{U}}} = -e\Delta V$. In addition, the difference between the corresponding LL indices provides a measure of $\Delta n$, $\Delta n = (N_{\mathrm{L}} - N_{\mathrm{U}})4eB/h$. From the values of $D$, $\Delta V$ and $\Delta n$ for a given LL crossing, exemplified by those shown in Figs. 4(a) and S8(b), we can extract $C_{\mathrm{GG}}$ from Eq. (3). The $C_{\mathrm{GG}}$ extracted from several LL crossings studied (see Fig. S9) are in good agreement with each other, with an average $C_{\mathrm{GG}}$ = (17.4 ± 0.5) μF/cm$^2$, and corresponding to $\varepsilon_{\mathrm{GG}}$ =6.7 $\varepsilon_0$ for $d_{\mathrm{GG}} = 0.34$ nm. The estimated $C_{\mathrm{GG}}$ is at least 2 (7) times of the value in large-$\theta$ tBLG (vacuum-filled parallel plate capacitor with inter-plate distance $d_{\mathrm{GG}}$ ) (see Figs. S10, S11 and Ref. [35]). Such a large $C_{\mathrm{GG}}$ is attributed to the reduced $v_{\mathrm{F}}$ in our small-$\theta$ tBLG. We find that the consideration of the effect of quantum capacitance will change the $C_{\mathrm{GG}}$ value by ~0.2 μF/cm$^2$, which is smaller than the uncertainty (~0.5 μF/cm$^2$) in the extracted $C_{\mathrm{GG}}$. Therefore, we ignore the effect of the quantum capacitance in tBLG. The enhancements of $C_{\mathrm{GG}}$ and $\varepsilon_{\mathrm{GG}}$ can also be explained qualitatively by the linear reduction of the Thomas-Fermi screening length with smaller $v_{\mathrm{F}}$ , $\lambda_{\mathrm{TF}} \propto v_{\mathrm{F}}/k_{\mathrm{F}}$ (here $k_{\mathrm{F}} = \sqrt{n\pi}$ is the Fermi momentum)[48], indicating a strong electronic screening in the small-$\theta$ tBLG.

**Conclusion**

In summary, we have performed temperature-dependent and magneto-transport studies on dual-gated tBLG samples with twist angle $\theta$~2° and encapsulated in h-BN. We have observed the transport features arising from the VHSs and SMGs in addition to the main DP. We have found that the resistivity measured between the CNP ($n$ ~0) and SMG exhibits a power-law behavior, $\sim T^{\beta}$. The extracted temperature exponent $\beta$ features a W-shaped carrier density dependence with two minima at the VHSs, indicating a distinct electron-phonon coupling for small-$\theta$ tBLG. From our experiment, we have measured the SMG gap size, which confirms its relatively large value as reported in a recent study[17]. We have also estimated the interlayer coupling strength, which may be useful for further studies on the origin of the large SMG gap. By measuring quantum oscillations at high magnetic fields, we have observed Berry phase transition from $\pi$ to $2\pi$ when increasing the carrier density and tuning the Fermi level across the VHS. Landau level crossings and Fermi velocity suppression observed at carrier densities below the VHS reveal strong interlayer coupling in the small-$\theta$ tBLG.

**Acknowledgements**

We acknowledge partial support of the work by NSF CMMI (Grant No. 1538360) and EFMA (Grant No. 1641101), and helpful discussions with M. Koshino.

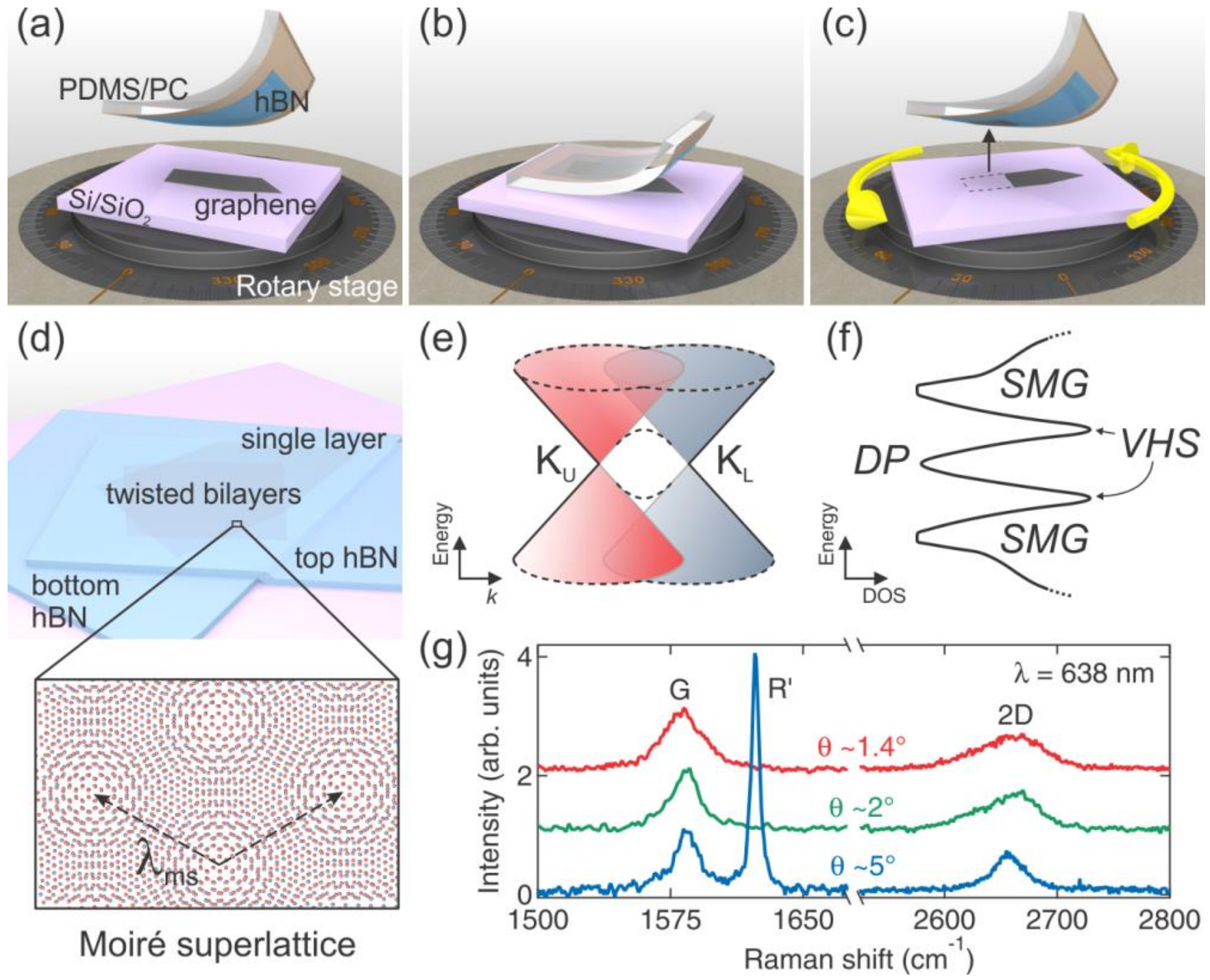

Figure 1. (a - d) Schematics of our technique for assembling twisted bilayer graphene (tBLG) encapsulated in hexagonal boron nitride (h-BN), with a controlled twist angle $\theta$ between the two monolayers (broken from the same piece of graphene single crystal). The inset below (d) shows the moiré superlattice of tBLG with a lattice constant $\lambda_{ms}$. Sketches of (e) tBLG band structure, showing Dirac cones at K valley of the upper and lower layers with a finite momentum separation, and of (f) its electronic density of states (DOS). The hybridization between the two graphene layers yields van Hove singularities (VHSs) and superlattice-induced mini-gaps (SMGs). The VHSs and SMGs are situated away from the charge neutrality point (CNP) and the main Dirac point (DP) of each Dirac cone. (g) Raman spectra of tBLG samples with $\theta$ of 1.4°, 2° and 5°. Spectra are individually normalized to the intensity of their respective G peak and are shifted vertically for clarity. Data were measured with 638 nm laser excitation.

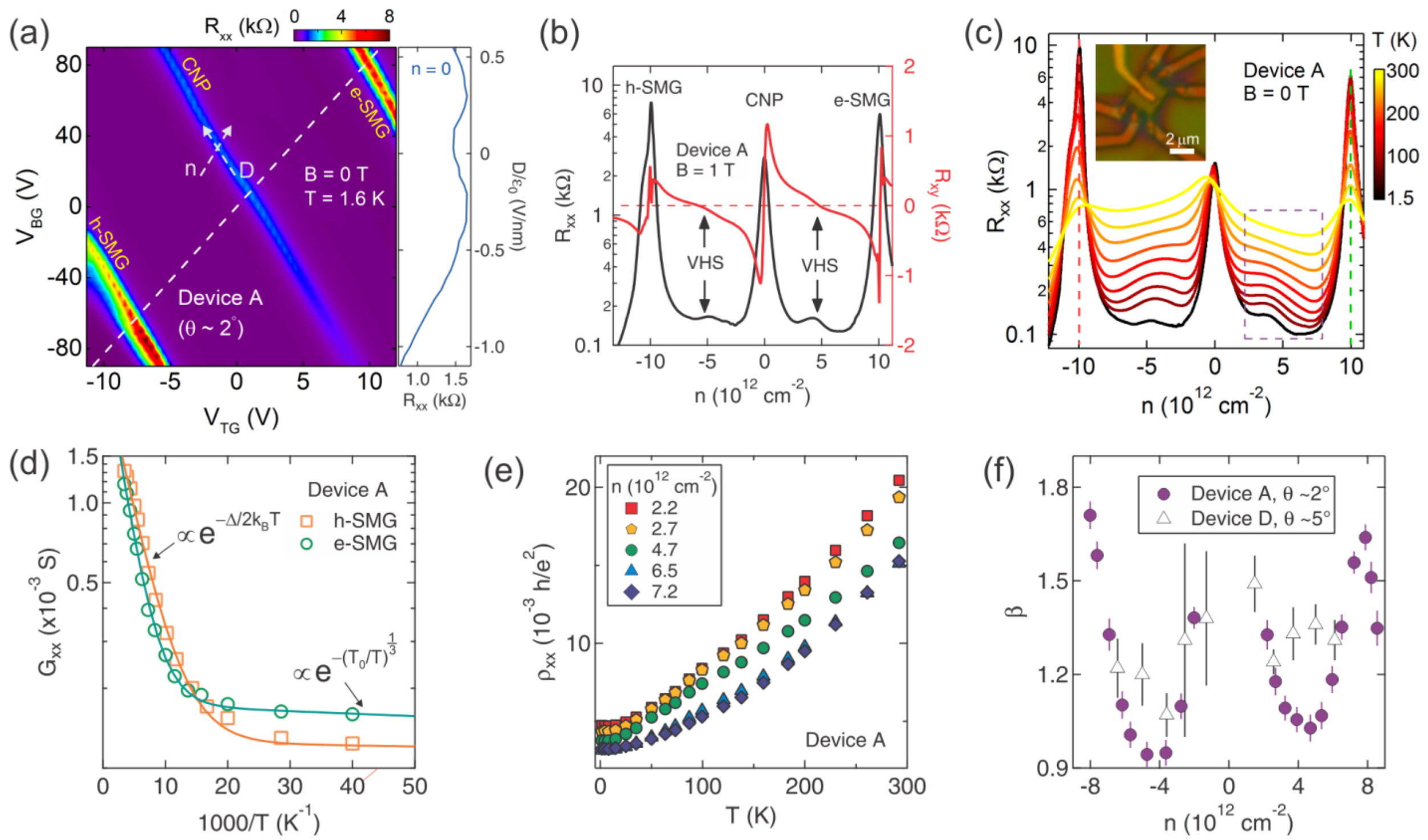

Figure 2. (a) Longitudinal resistance ($R_{xx}$) (color scale) as a function of top-gate voltage ($V_{TG}$) and back-gate voltage ($V_{BG}$) for tBLG Device A with $\theta$~2° measured at zero magnetic field ($B$ = 0 T) and temperature ($T$) of 1.6 K. There are two dashed arrows, indicating the axes of $n$ and $D$ (total carrier density and average displacement field applied normal to the graphene layers; see also the definition in the main text). Along the $n$-axis, $D$ = 0 when there is no interlayer voltage difference (the Dirac cones in the two layers are aligned in energy; see Fig. 1(e)), while along the $D$-axis, $n$ = 0 when the total carrier density in the two layers is zero. The inset shows $R_{xx}$ extracted along $n$ = 0 (along the central blue stripe in the main panel) versus $D/\varepsilon_0$. (b) $R_{xx}$ (in log-scale) and Hall resistance ($R_{xy}$) of Device A measured as functions of $n$ along the dashed line in (a) by tuning $V_{TG}$ and $V_{BG}$ simultaneously at $B$ = 1 T and $T$ = 1.6 K. Sign reversal in the $R_{xy}$ at CNP, VHSs, and SMGs indicates a change in charge carrier type (from electron to hole or vice versa). The two shallow resistance peaks in $R_{xx}$ corresponds to the two VHSs, where $R_{xy}$ also crosses zero. (c) $R_{xx}$ (at $B$ = 0 T) of Device A as a function of $n$ along the dashed line in (a) at various $T$, showing the insulating behavior around $n = n_s = \pm 9.9 \times 10^{12}$ cm$^{-2}$, from which the twist angle $\theta$ is estimated. The inset shows an optical image of device A. (d) Arrhenius plot of the conductance ($G_{xx} = 1/R_{xx}$) extracted at $n_s$ for the SMGs. The solid lines are fits to $G_{xx} = G_{TA}\cdot\exp(-\Delta/2k_BT) + G_{VRH}\cdot\exp[-(T_0/T)^{1/3}]$ (see the main text for details). The activation gap ($\Delta$) is ~65 meV and ~45 meV for the electron- and hole-side mini-gaps, respectively. (e) $T$-dependence of resistivity ($\rho_{xx}$) for $n$ from 2.2 to 7.2 × $10^{12}$ cm$^{-2}$, in the range marked by the dashed rectangle in (c), exhibiting metallic behavior ($d\rho_{xx}/dT > 0$). The $T$-dependence below ~150 K can be fitted to $\Delta\rho_{xx}(T,n) = \rho_{xx}(T,n) - \rho_0(n) = \alpha T^{\beta}$, attributed to acoustic phonon scattering. (f) Fitted $T$-exponent ($\beta$) as a function of $n$ for Devices A ($\theta$~2°) and D ($\theta$ ~5°).

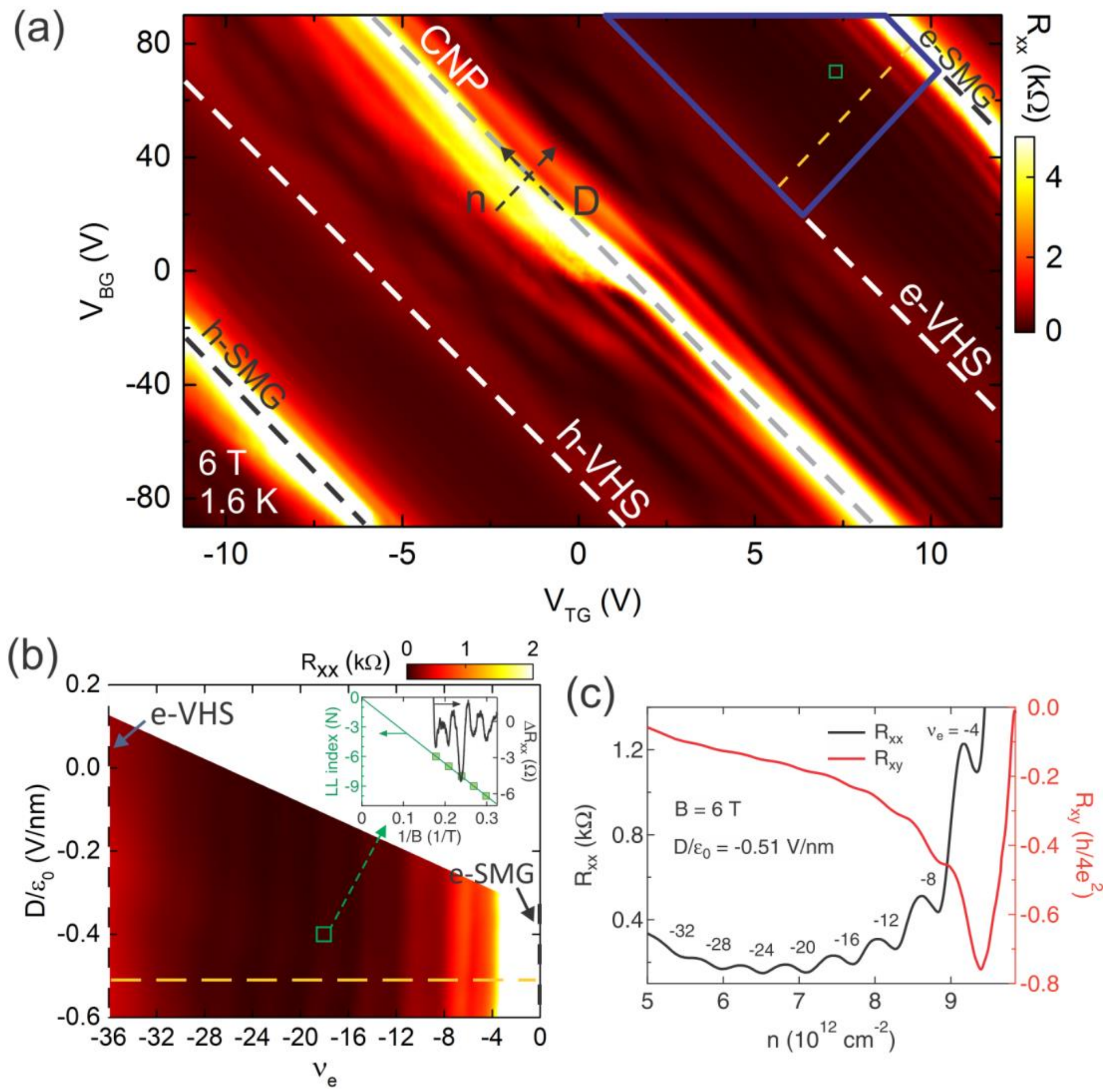

Figure 3. (a) Longitudinal resistance ($R_{xx}$) (color scale) as a function of $V_{TG}$ and $V_{BG}$ for Device A, measured at $B$ = 6 T and $T$ = 1.6 K. For carrier density $n$ between the two VHSs, we observe crossing of two sets of Landau levels (LL) when the layer degeneracy is broken by applying $D$. In contrast, only one set of LLs (manifested as lines parallel to the $D$-axis) are observed for $n$ beyond those of the VHS in the electron- or hole-side of CNP. (b) Zoomed-in color scale plot of the $R_{xx}$ (from the region bounded by blue solid lines in (a), between the VHS and SMG in the electron-side of CNP) as a function of $D/\varepsilon_0$ and filling factor ($\nu_e$, measured from the e-SMG), showing developing quantum Hall (QH) states (occurring in steps of 4 in $\nu_e$). The inset shows the assigned LL index ($N$) and corresponding Shubnikov-de Haas (SdH) oscillations in $\Delta R_{xx}$ ($R_{xx}$ with background subtracted) versus $1/B$, taken at fixed gate voltages (marked by the green open square in (a, b)) with $D/\varepsilon_0 \sim$ -0.4 V/nm and $n - n_{e\text{-}SMG}$ = $-3.2 \times 10^{12}$ cm$^{-2}$ (measured from the e-SMG; the negative sign represents hole-like carriers). The solid line is a linear fit with $N$ axis intercept -0.07 $\pm$ 0.05, indicating zero Berry phase (different from the massless charge carriers in monolayer). (c) $R_{xx}$ and $R_{xy}$ versus $n$ at $D/\varepsilon_0$ = -0.51 V/nm, measured along the orange dashed line in (a, b).

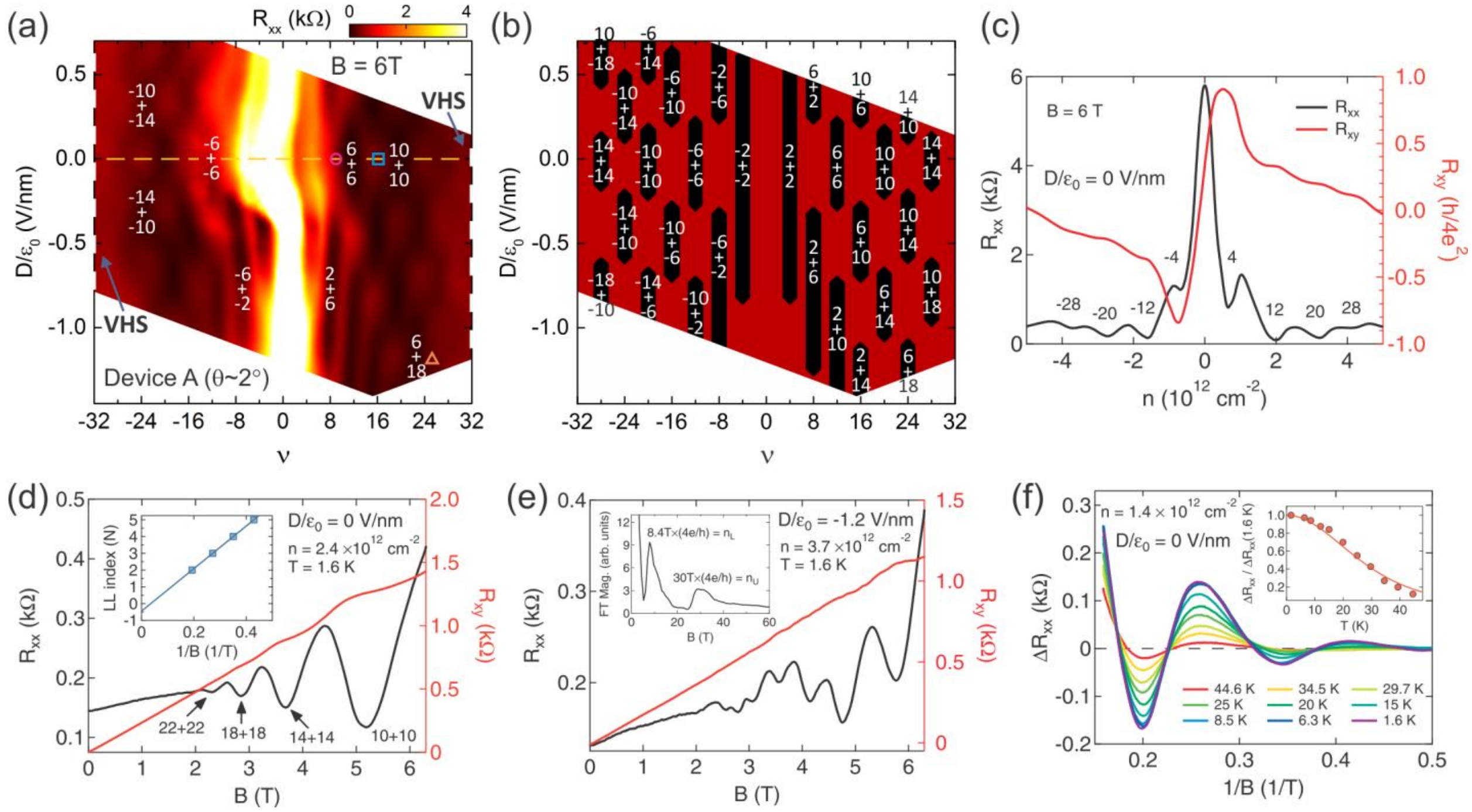

Figure 4. (a) Close-ups of the $R_{xx}$ (color scale) between the two VHSs in Fig. 3(a) as a function of $D/\varepsilon_0$ and $\nu$, measured at 6 T and 1.6 K. (b) Schematic (adapted from Ref. [35]) of expected QH states (regions in black) with corresponding filling factor combination ($\nu = \nu_L + \nu_U$) in the measured range in (a). (c) $R_{xx}$ and $R_{xy}$ at $D$ = 0, measured along the orange dashed line in (a), as functions of $n$. The $\nu$ associated with the minima in $R_{xx}$ are $\pm$4, $\pm$12, $\pm$20 and $\pm$28 (indicating 8-fold degenerate LL). (d) $R_{xx}$ and $R_{xy}$ as functions of $B$ measured at $D$ = 0 and $n = 2.4 \times 10^{12}$ cm$^{-2}$ (marked by the blue open square in (a)), showing SdH oscillations from two decoupled graphene monolayers with the same carrier density ($n$/2). The inset displays the assigned LL index ($N$) plotted against 1/$B$. The solid line is a linear fit with $N$ axis intercept -0.49 $\pm$ 0.02, which indicates π Berry phase for massless Dirac fermions. (e) $R_{xx}$ and $R_{xy}$ versus $B$ measured at $D/\varepsilon_0$ = -1.2 V/nm and $n = 3.7 \times 10^{12}$ cm$^{-2}$ (marked by the orange open triangle in (a)). Here the oscillations arise from the two decoupled monolayers, where the layer degeneracy in the LLs (and layer density) has been lifted by $D \neq 0$. The inset shows the magnitude of Fourier transform of $R_{xx}$(1/B). The two peaks at 8.4 T and 30 T correspond to the two different layer densities $n_U$ and $n_L$, respectively. (f) Temperature dependence of the SdH oscillations in $\Delta R_{xx}$ ($R_{xx}$ with background subtracted) at $n = 1.4 \times 10^{12}$ cm$^{-2}$ and $D$ = 0 (denoted by the pink open circle in (a)). The inset presents the temperature dependence of the normalized amplitude of $\Delta R_{xx}$ for the oscillation at 5 T ($\nu$ = 6+6 QH state). The solid line is a fit to the Lifshitz–Kosevich formula, yielding the electron effective mass ($m^*$ ~0.029$m_e$) and Fermi velocity ($v_F$ ~0.58 $\times$ $10^6$ ms$^{-1}$).

# Supplemental Material

Ting-Fung Chung,[1] Yang Xu,[1] and Yong P. Chen[1,2,3,†]

[1] *Department of Physics and Astronomy and Birck Nanotechnology Center, Purdue University, West Lafayette, Indiana 47907, USA*

[2] *School of Electrical and Computer Engineering and Purdue Quantum Center, Purdue University, West Lafayette, Indiana 47907, USA*

[3] *WPI-AIMR International Research Center on Materials Science, Tohoku University, Sendai 980-8577, Japan*

## I. Sample assembly and fabrication

We assembled twisted bilayer graphene (tBLG) using the dry-transfer method[1,2], which allows us to control the twist angle ($\theta$) of tBLG. The process started with exfoliating graphene (Kish Graphite from Covalent Materials Corp.) and h-BN (bulk hexagonal boron nitride from HQ graphene) on clean silicon (Si) substrates with a ~290 nm-thick thermal oxide ($SiO_2$) overlayer. We picked up the top h-BN with a PPC spin-coated PDMS/glass stamp (here PPC is polypropylene carbonate and PDMS is poly-dimethylsiloxane). With the aid of a micro-manipulator and rotary stage, we achieved angle alignment of tBLG by breaking and picking up a piece from the same large single crystal graphene flake (see schematics in Fig. 1 in the main text). For Device A, C and D, the h-BN/graphene/graphene stack was released to the bottom h-BN flake exfoliated on a different $SiO_2$/Si substrate. In Device B, the stack placed on an h-BN flake at the bottom was transferred onto a metal gate fabricated on a $SiO_2$/Si substrate. The thickness of top and bottom h-BN used varies from ~12 to ~45 nm, as measured by atomic force microscopy (AFM).

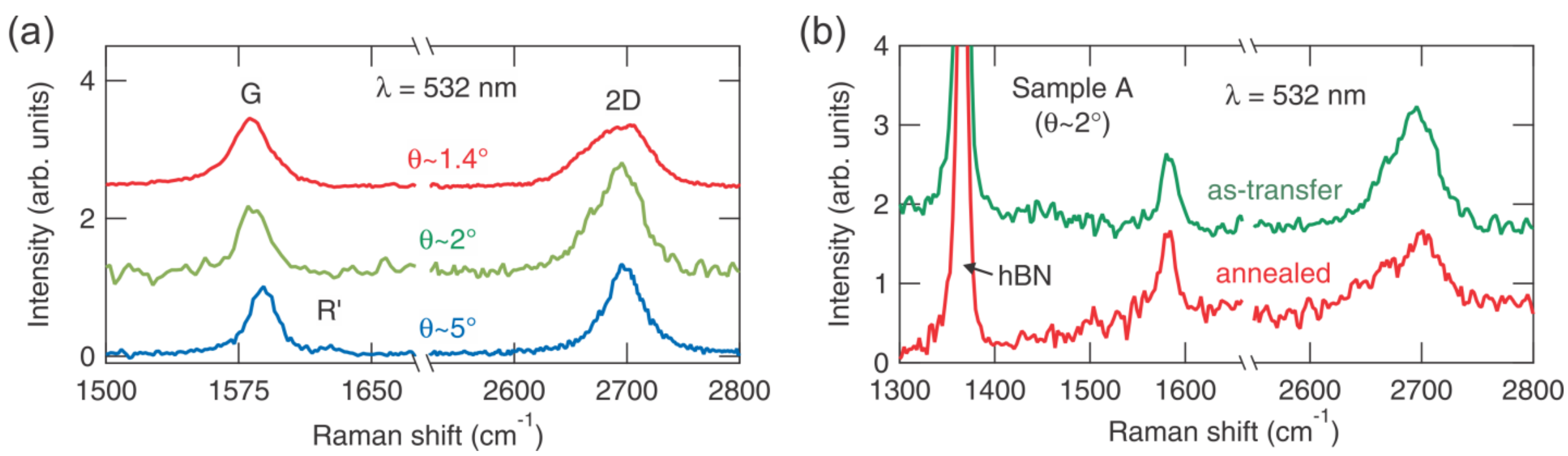

Figure S1. (a) Raman spectra of tBLG with $\theta$ of 1.4°, 2° and 5°, measured with a 532 nm laser excitation. The Raman 2D peak of tBLG broadens asymmetrically with decreasing $\theta$. The intensity of the R' peak seen in the sample at $\theta$ ~5° is largely suppressed because of off-resonance excitation (in contrast to Fig.1 in the main text, measured with a 638 nm laser)[3]. (b) Comparison of the Raman spectra of Sample A ($\theta$ ~2°) taken before and after thermal annealing. The giant peak at ~1366 $cm^{-1}$ stems from h-BN layers. All spectra are individually normalized to the intensity of their respective G peak and are shifted vertically for clarity.

After the stack (h-BN/tBLG/h-BN) was made, we used Raman spectroscopy to characterize and estimate $\theta$ (Ref. [4]). For the samples with $\theta \leq 2°$, we further confirmed their $\theta$ using field effect measurements (discussed in the main text). We annealed the samples in forming gas (5% $H_2$ and 95% Ar) at 500 °C for 30 minutes to release trapped bubbles. The ramping and cooling rates of annealing were slow as graphene layers could easily rotate at high heating rates[5]. Figure S1(b) shows a comparison of the Raman spectra of Sample A ($\theta$ ~2°) taken with a 532 nm excitation laser before and after annealing. After annealing, we observe a rising background at higher wavenumbers, which could be attributed to the emissions from increasing h-BN defects[6] and from organic residues introduced by the dry-transfer[7].

We also see a slight reduction in the 2D band intensity, which becomes comparable to the intensity of the G band after annealing. We used AFM and optical microscopy to find bubble-free regions for the device fabrication.

For the device fabrication, we used standard electron-beam lithography (EBL) and reactive ion etching (RIE) with multiple steps. First, we made the top-gate (5/50 nm of Cr/Au). Second, we used a negative photoresist Ma-N 2403 together with the metal top-gate as an etch mask to define the channel and leads of the device, followed by dry etching with Ar/$SF_6$/$O_2$ mixture. We then used another PMMA mask to define contacts. The edge contact scheme[1] was used and gentle $O_2$ plasma ashing (cleaning) was further applied before metal deposition (10/60–70 nm of Cr/Au). To connect the top-gate and avoid its short-circuiting with the channel, we covered the device with another thin h-BN flake. Finally, we created a via connecting the metal top-gate. The channels defined by the top-gate metal and dry etching are ~2.1 μm long and ~2.6 μm wide for Device A, ~4.5 μm long and ~1 μm wide for Device B, ~1.2 μm long and ~0.9 μm wide for Device C, and ~2.2 μm long and ~1.5 μm wide for Device D, respectively.

## II. Dual-gated field effect measurement at zero magnetic field

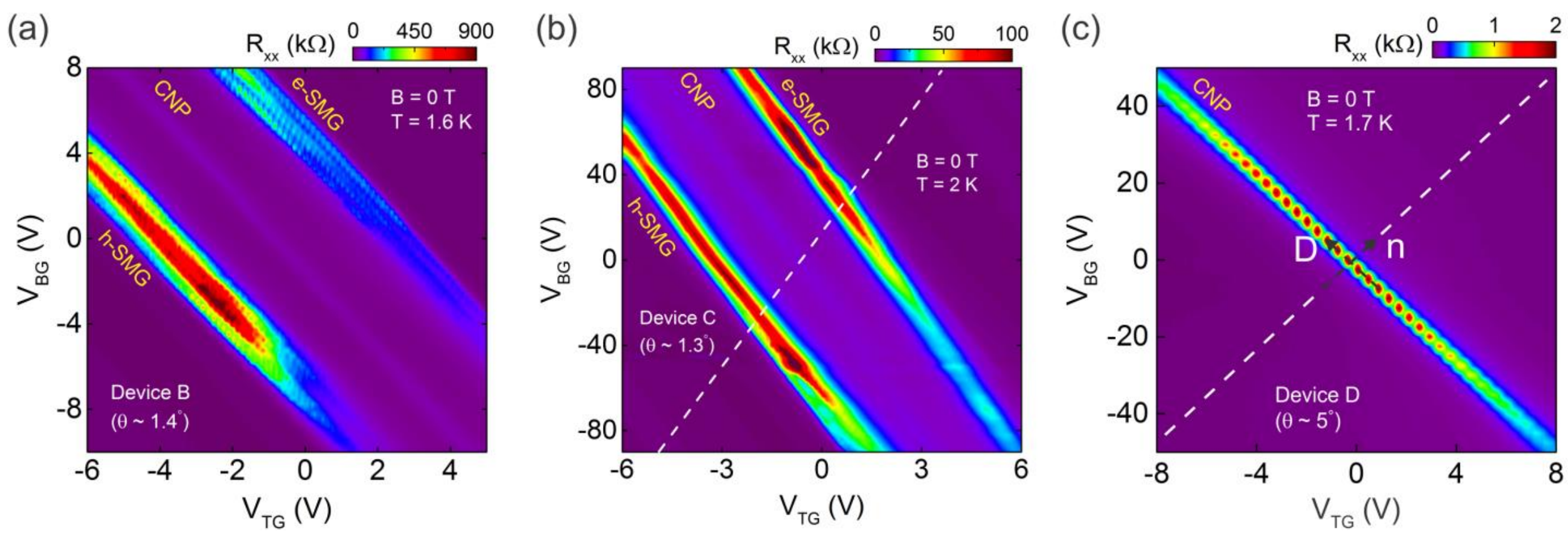

Figure S2. Longitudinal resistance ($R_{xx}$) (color scale) as a function of top-gate voltage ($V_{TG}$) and back-gate voltage ($V_{BG}$) for tBLG Device B with $\theta$~1.4° (a), Device C with $\theta$~1.3° (b), and Device D with $\theta$~5° (c). All measurements were performed at $B$ = 0 T and $T$ ~2 K.

## III. Temperature dependence and energy gap of SMG

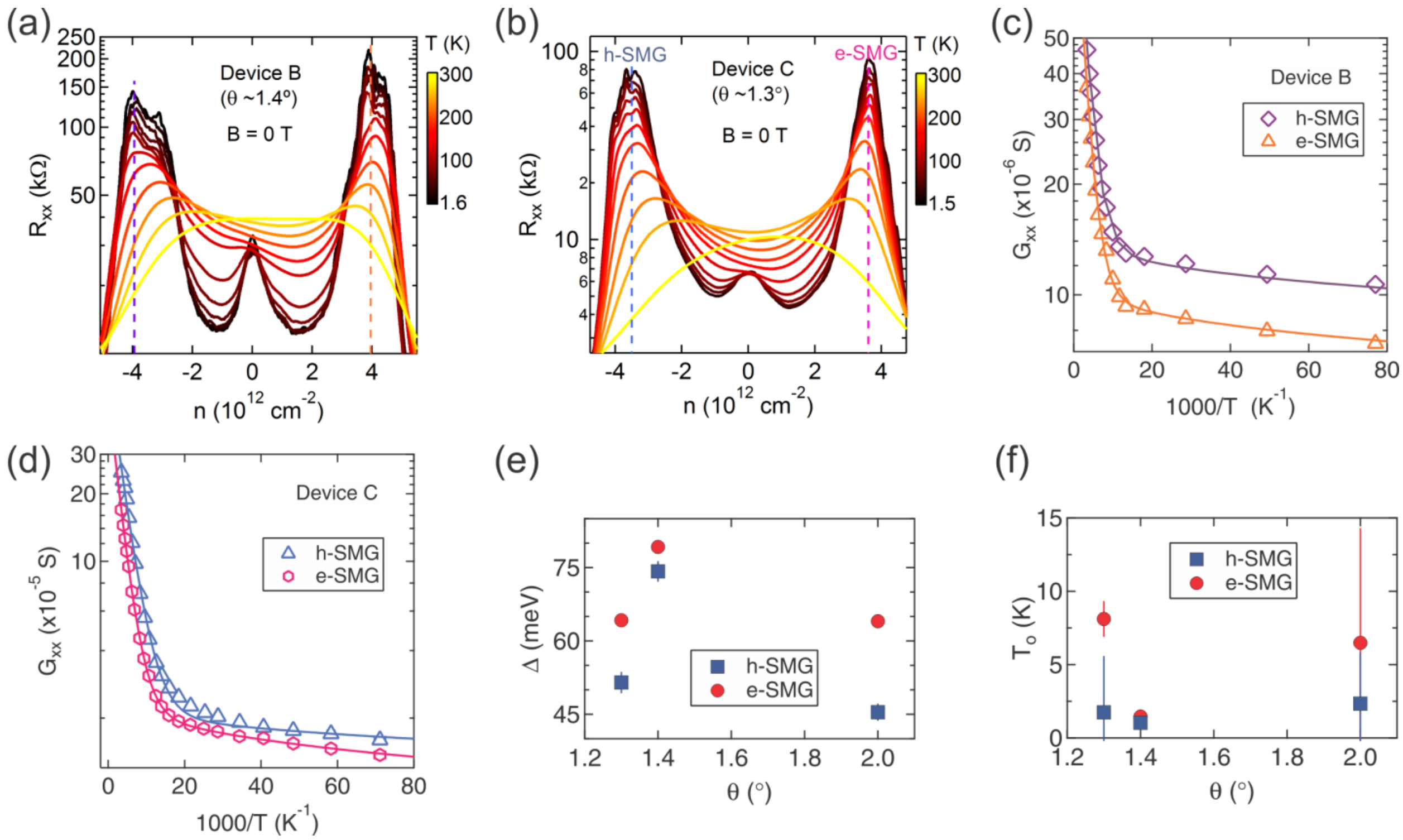

Figure S3. (a) $R_{xx}$ (at $B$ = 0 T) of Device B versus carrier density ($n$), measured by sweeping $V_{BG}$ at different $T$. The SMGs in Device B are located at $n_s \approx \pm 4 \times 10^{12}$ cm$^{-2}$. (b) $R_{xx}$ (at $B$ = 0 T) of Device C versus $n$ along the dashed line in Fig. S2(b) at various $T$, manifesting the SMGs at $n_s \approx \pm 3.6 \times 10^{12}$ cm$^{-2}$. Arrhenius plot of the conductance ($G_{xx} = 1/R_{xx}$) extracted at $\pm n_s$ for the SMGs in Device B (c) and Device C (d). The solid lines represent best fits to $G_{xx} = G_{TA}\exp(-\Delta/2k_B T) + G_{VRH}\exp(-(T_o/T)^{1/3})$, which is the same equation discussed in the main text. We can obtain two fitting parameters: the activation gap (Δ) and the characteristic temperature ($T_o$) for VRH. The results from three different tBLG devices with small-$\theta$ are presented in (e) and (f) as a function of $\theta$. The date Δ in (e) exhibits a trend that first increases with decreasing $\theta$ down to $\theta$ ~1.5° and then decreases for further smaller $\theta$, qualitatively consistent with predictions based on strong electronic correlations in the moiré superlattices in the very small-$\theta$ regime.[8,9] Further studies with various $\theta$ is needed to verify the trend between Δ and $\theta$.

IV. Comparison of the temperature-dependent resistivity at different carrier densities

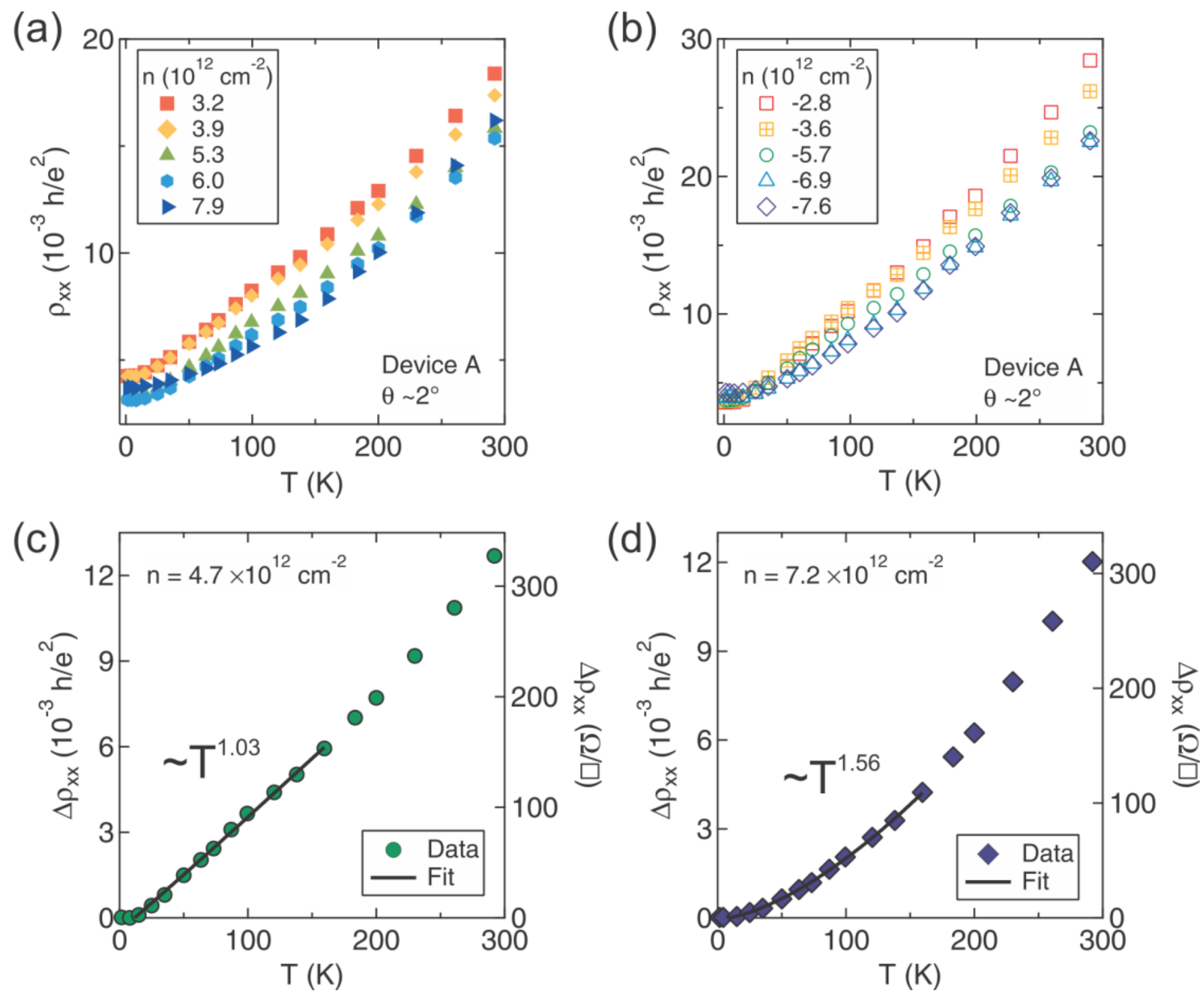

Figure S4. $T$-dependence of resistivity ($\rho_{xx}$ or sheet resistance) and representative fits for Device A ($\theta \sim 2°$) at various $n$. (a) For $n$ spanning from $3.2 \times 10^{12}$ to $7.9 \times 10^{12}$ cm$^{-2}$, extracted from the range marked by the dashed rectangle in Fig. 2(c) of the main text. (b) For $n$ ranging from $-2.8 \times 10^{12}$ to $-7.6 \times 10^{12}$ cm$^{-2}$, extracted from the same device in Fig. 2(c). The difference between measured resistivity and residual resistivity ($\Delta\rho_{xx} = \rho_{xx}(n,T) - \rho_0(n)$) follows a power law, $\Delta\rho_{xx} = \alpha T^{\beta}$. Representative fits to the power law at $T < 160$ K (without accounting for high energy phonons) for $n = 4.7 \times 10^{12}$ cm$^{-2}$ (c) and for $n = 7.2 \times 10^{12}$ cm$^{-2}$ (d). The fits (solid lines) yield $T$-exponent $\beta = (1.03 \pm 0.04)$ and $(1.56 \pm 0.04)$ for $n = 4.7 \times 10^{12}$ cm$^{-2}$ and $7.2 \times 10^{12}$ cm$^{-2}$, respectively.

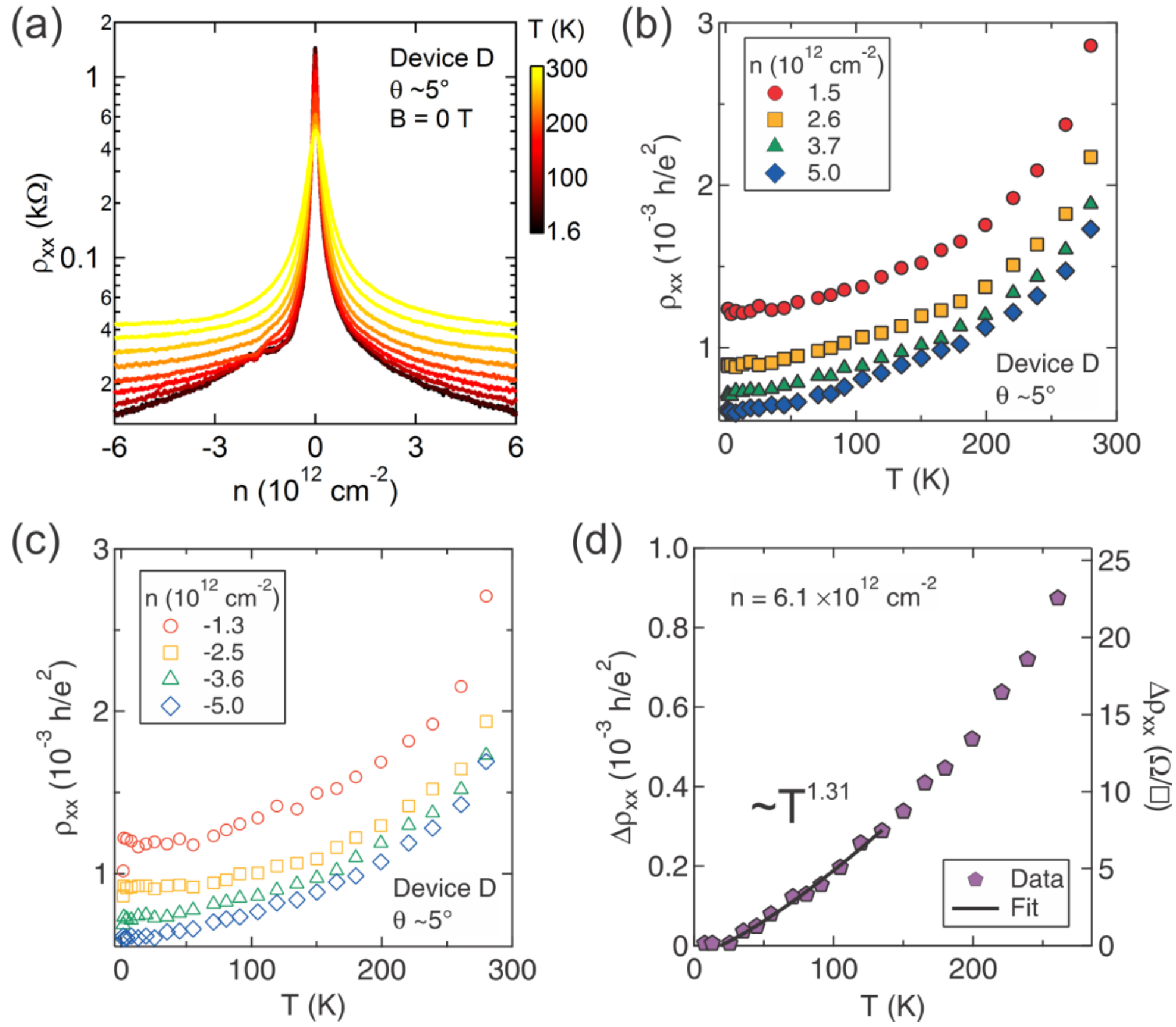

Figure S5. $T$-dependent field effect ($\rho_{xx}$ versus $n$), $T$-dependent $\rho_{xx}$ at various $n$, and a representative fit to the $T$-dependent $\rho_{xx}$ for Device D ($\theta$ ~5°). The device has a Hall mobility ~137,000 $cm^2V^{-1}s^{-1}$ for $n \approx 1.5 \times 10^{12}$ $cm^{-2}$ at $T$ = 1.6 K. (a) Field effect curves along the dashed line in Fig. S2(c) at several representative $T$. $T$-dependence of $\rho_{xx}$ (b) for $n = 1.5 \times 10^{12}$ to $5 \times 10^{12}$ $cm^{-2}$ (e-doped), and (c) for $n = -1.3 \times 10^{12}$ to $-5 \times 10^{12}$ $cm^{-2}$ (p-doped). Data are extracted from the range of $n$ in (a). A representative fit for $n = 6.1 \times 10^{12}$ $cm^{-2}$ is presented in (d). The solid line is a fit to $\Delta\rho_{xx} = \alpha T^{\beta}$ at $T < 150$ K, giving $\beta = 1.31 \pm 0.07$.

V. Measurement of Fermi velocity

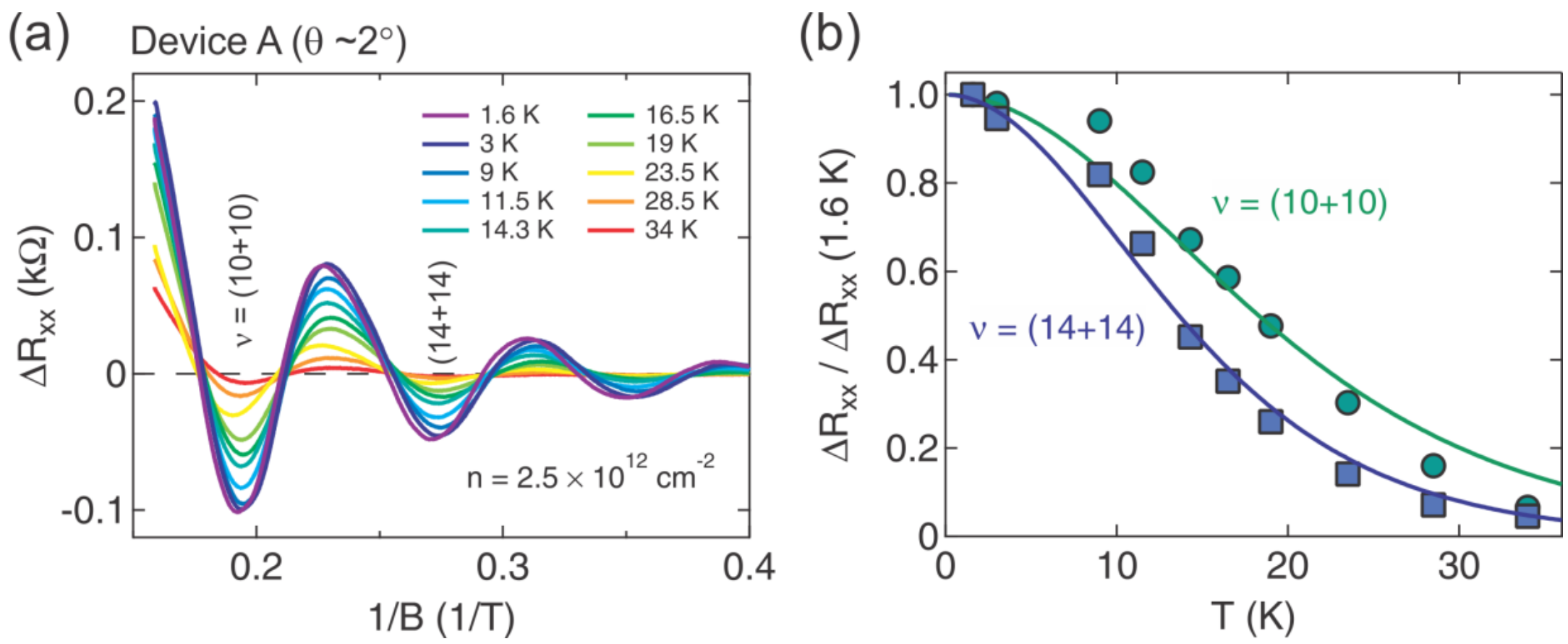

Figure S6. (a) *T*-dependence of the SdH oscillations in $\Delta R_{xx}$ ($R_{xx}$ with background subtracted) for Device A ($\theta$~2°) at $n$ = 2.5 × 10$^{12}$ cm$^{-2}$. The average displacement field ($D$) was close to zero in the measurement. (b) *T*-dependence of the scaled oscillation amplitude ($\Delta R_{xx}/\Delta R_{xx}(T$ = 1.6 K)) at $B$ = 5.17 T (total filling factor, $\nu = \nu_L + \nu_U$ = 10+10) and 3.68 T ($\nu$ = 14+14), giving the carrier effective masses ($m$*) of 0.042$m_e$ and 0.040$m_e$, respectively. Their respective Fermi velocities ($v_F$) are 0.54 × 10$^6$ ms$^{-1}$ and 0.57 × 10$^6$ ms$^{-1}$. Symbols are experimental data and solid lines are fits to the Lifshitz-Kosevich formula[10].

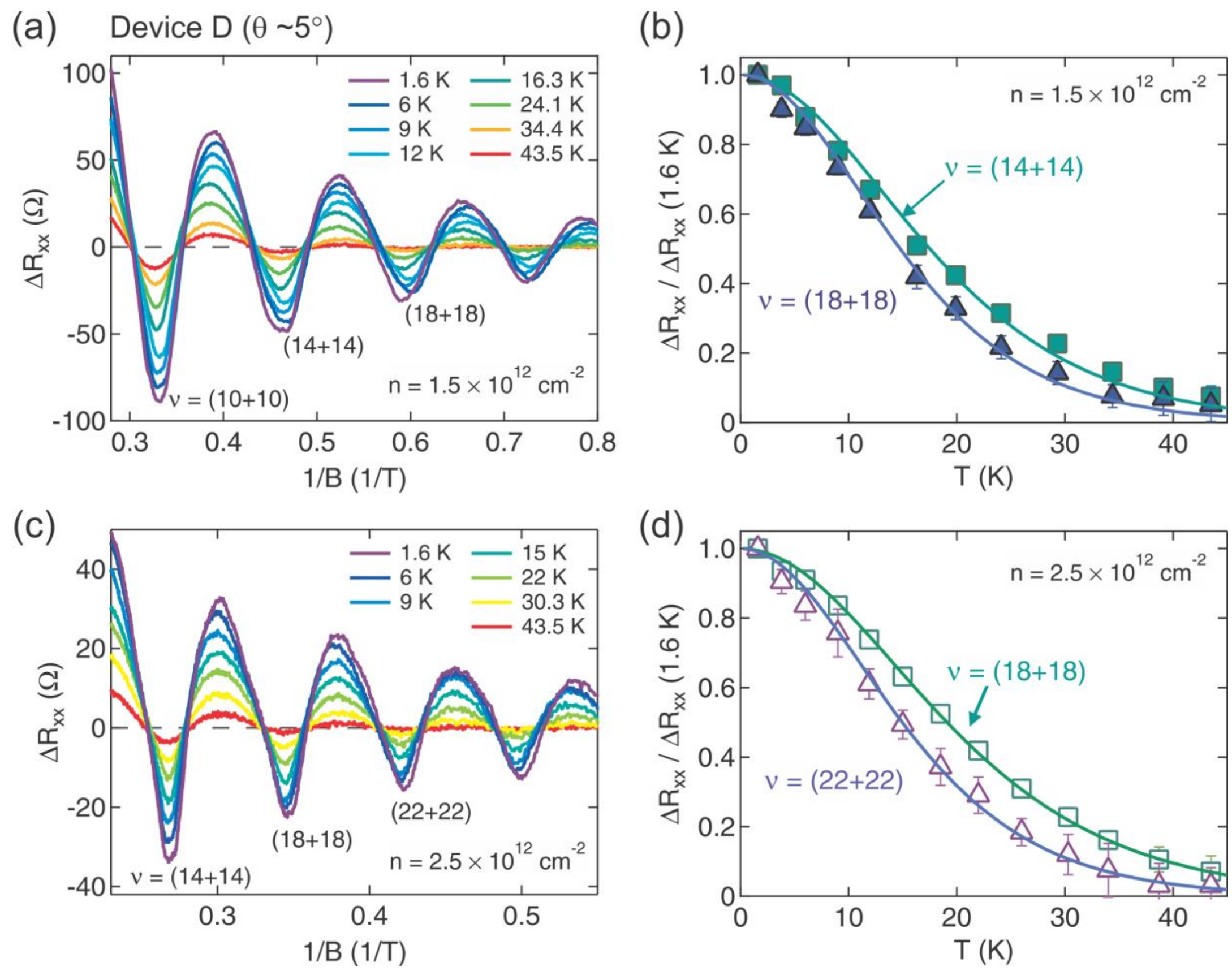

Figure S7. $T$-dependence of the SdH oscillations in $\Delta R_{xx}$ ($R_{xx}$ with background subtracted) for Device D ($\theta$ ~5°) at $n = 1.5 \times 10^{12}$ cm$^{-2}$ (a) and $2.5 \times 10^{12}$ cm$^{-2}$ (c). $D \sim 0$ were maintained in all measurements. (b) $T$-dependence of the normalized oscillation amplitude at $B$ = 2.16 T ($\nu$ = 14+14) and 1.69 T ($\nu$ = 18+18) for $n = 1.5 \times 10^{12}$ cm$^{-2}$, giving $m^* = 0.018m_e$ and $0.017m_e$, and $v_F$ = (0.96 and 1.04) × 10$^6$ ms$^{-1}$, respectively. (d) Same as in (b) at $B$ = 2.89 T ($\nu$ = 18+18) and 2.36 T ($\nu$ = 22+22) for $n = 2.5 \times 10^{12}$ cm$^{-2}$. Fits yield $m^* = 0.022m_e$ and $v_F = 1.03 \times 10^6$ ms$^{-1}$ for $\nu$ = 18+18 QH state, and $m^* = 0.023m_e$ and $v_F = 0.98 \times 10^6$ ms$^{-1}$ for $\nu$ = 22+22 QH state. The solid lines are fits to the Lifshitz-Kosevich formula[10]. Note that the Lifshitz-Kosevich formula does not provide a good fit for the normalized oscillation amplitude data of $\nu$ = 10+10 QH state for $n = 1.5 \times 10^{12}$ cm$^{-2}$ in (a) and of $\nu$ = 14+14 QH state for $n = 2.5 \times 10^{12}$ cm$^{-2}$ in (c). Those data are not shown in (b,d).

VI. Analysis of Landau level crossing

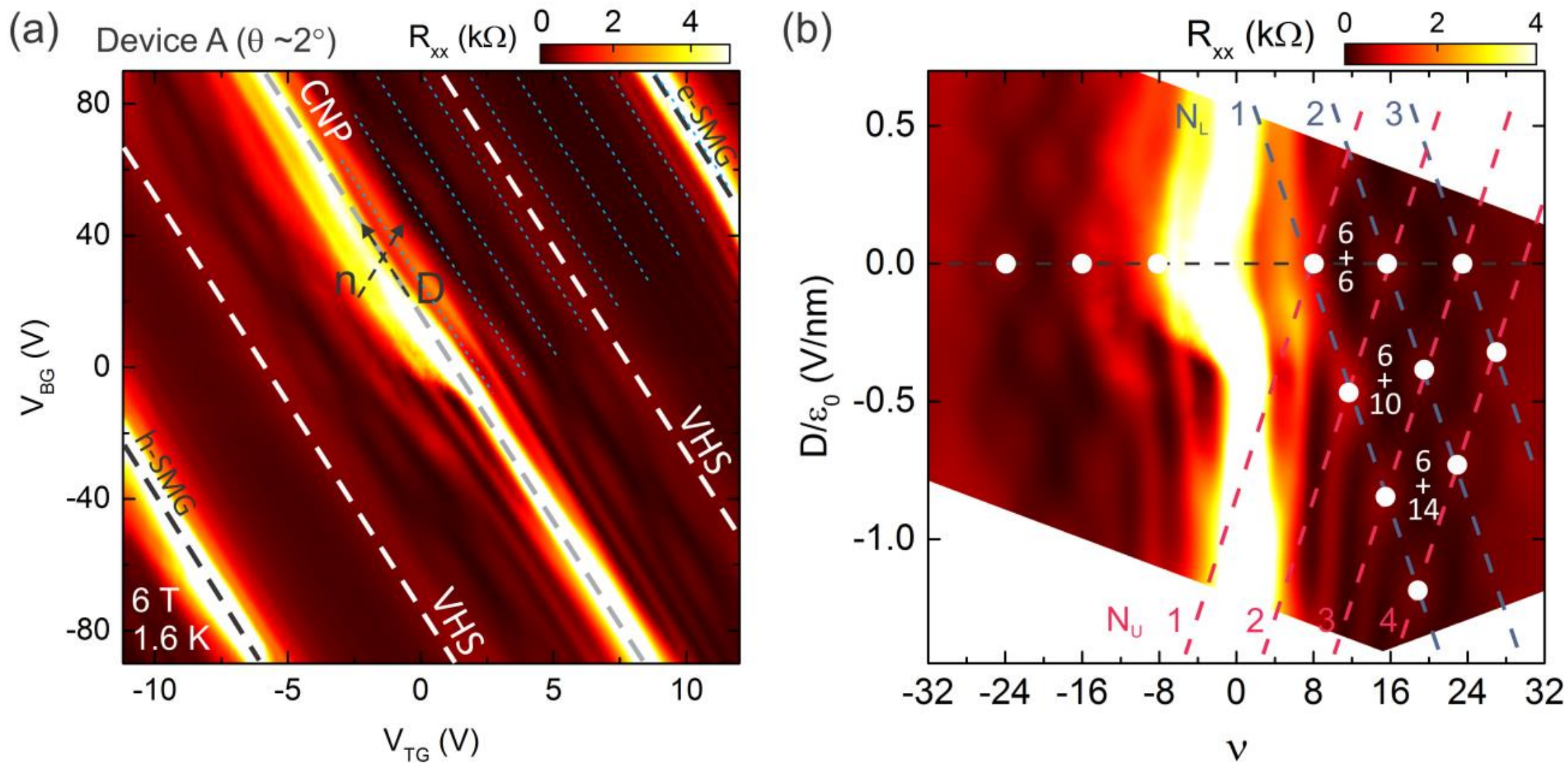

Figure S8. (a) $R_{xx}$ (color scale) as a function of $V_{TG}$ and $V_{BG}$ for Device A, measured at $B$ = 6 T and $T$ = 1.6 K (same as Fig. 3(a) in the main text). Blue dotted lines (parallel to those corresponding to CNP, VHS, and SMG) indicate QH states with $\nu = 8(N + 1/2)$, where the degeneracy factor is 8 (2-layer, 2-spin and 2-valley) and Landau level (LL) index ($N$) = 0, ±1, ±2, …. (b) Zoomed color scale plot of the $R_{xx}$ (same data as in Fig. 4(a) in the main text) as a function of $D/\varepsilon_0$ and $\nu$ at 6 T and 1.6 K. White dots indicate the LL crossings where the interlayer capacitance per unit area ($C_{GG}$) and the interlayer dielectric constant ($\varepsilon_{GG}$) are extracted. Red and blue dashed lines, as guides to the eye, denote the LL index of the upper ($N_U$) and lower ($N_L$) graphene layers.

Figure S8 shows $R_{xx}$ versus $V_{TG}$ and $V_{BG}$ for Device A, measured at constant $B$ = 6 T, revealing crossing of two sets of Landau level (LL crossings) for $n$ between e- and h-VHSs, while only one set of LL for $n$ beyond the VHSs is seen. We take a closer look at the VHS energy ($E_{VHS}$) in this sample. We indicate the QH states in the e-doped side of CNP with blue dotted lines parallel to the CNP ($n$ = 0). These dotted lines are equally-spaced, passing through points of equal filling factors $\nu = nh/eB = 4(N + 1/2) =$ 2, 6, 10 … in the two layers, where $h$ is the Planck's constant, $e = 1.602 \times 10^{-19}$ C is the elementary charge, the degeneracy factor is 4 (2-layer, 2-spin for one layer), and $N$ is the LL index. As illustrated in Fig. S8(a), e-VHS is located at the middle between the lines with $\nu$ = 14 ($N$ = 3) and 18 ($N$ = 4). Thus, we consider the carrier density and the LL energy at e-VHS similar to those for $N$ ~3.5 at $B$ = 6 T.

In the decoupled regime ($n$ below VHS), the LL spectrum of each graphene layer in the tBLG can be described as massless Dirac fermions with a reduced $v_F$ (Ref. [11]). The LL spectrum follows the sequence, $E_{\mathrm{N}} = \mathrm{sgn}(N)\sqrt{2e\hbar v_{\mathrm{F}}^2 B|N|}$ . With the input of $v_{\mathrm{F}}$ ~0.58 × $10^6$ $\mathrm{ms}^{-1}$ (measured from the *T*-dependence of SdH oscillations), we can estimate $E_{\mathrm{VHS}} = E_{\mathrm{N=3.5}}$ ~95 meV, which is consistent with a recent STM result[12]. Note that it is difficult to estimate the e-SMG energy ( $E_{\mathrm{SMG}}$ , the energy difference between CNP and e-SMG) with the same method because $v_{\mathrm{F}}$ and the LL energy spectrum in the coupled regime with quadratic band near SMG[8,13] of tBLG are unknown.

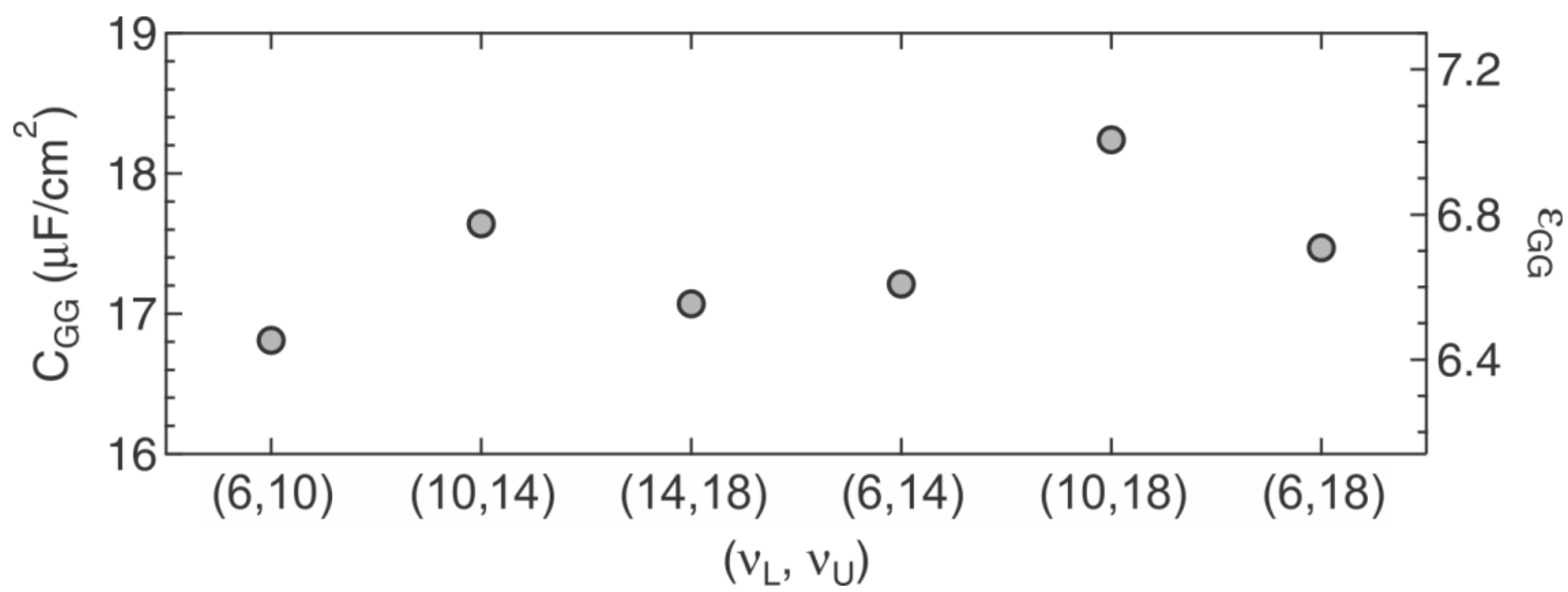

Figure S9. $C_{\mathrm{GG}}$ and $\varepsilon_{\mathrm{GG}} = d_{\mathrm{GG}} C_{\mathrm{GG}} / \varepsilon_0$ extracted from several LL crossings (with filling factors ( $\nu_{\mathrm{L}}$ , $\nu_{\mathrm{U}}$ ) from the lower and upper layers) in Device A (see Fig. S8), showing consistent values for each LL crossing. We obtain an average $C_{\mathrm{GG}}$ = (17.4 ± 0.5) μF/cm$^2$, corresponding to $\varepsilon_{\mathrm{GG}} \approx 6.7\,\varepsilon_0$ for $d_{\mathrm{GG}}$ = 0.34 nm. The error in $C_{\mathrm{GG}}$ denotes one standard deviation.

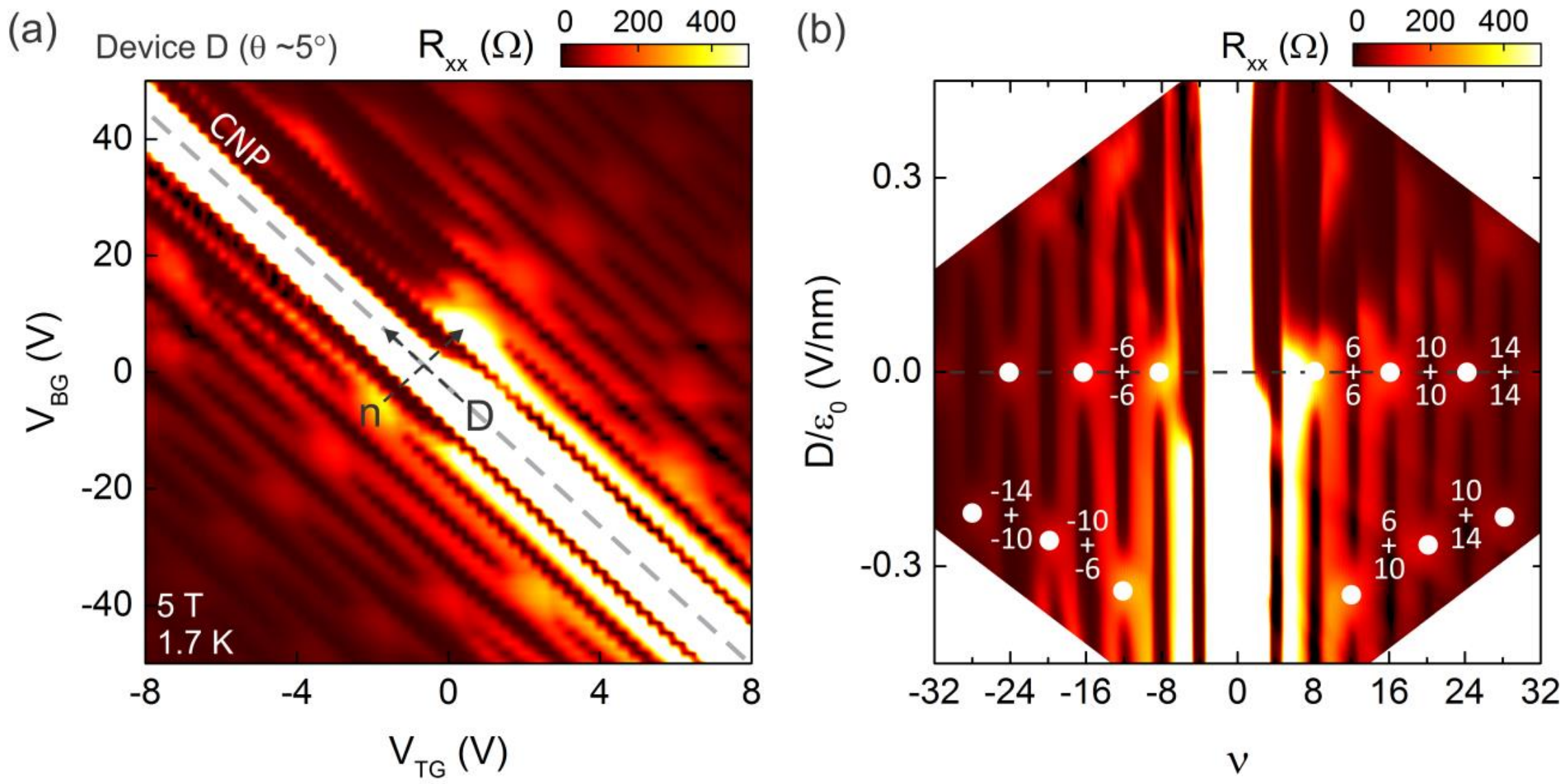

Figure S10. (a) $R_{xx}$ (color scale) as a function of $V_{TG}$ and $V_{BG}$ for Device D, measured at $B$ = 5 T and $T$ = 1.7 K, displaying crossings of two sets of LL over wide ranges of $n$ and $D$. The LL crossings observed are consistent with prior reports in large-$\theta$ tBLG[14,15]. (b) $R_{xx}$ (near the CNP shown in (a)) as a function of $D/\varepsilon_0$ and $\nu$ at 5 T and 1.7 K. White dots mark the crossings of two LLs. The expected filling factor combinations $(\nu_L + \nu_U)$ are shown in corresponding QH states (regions in dark).

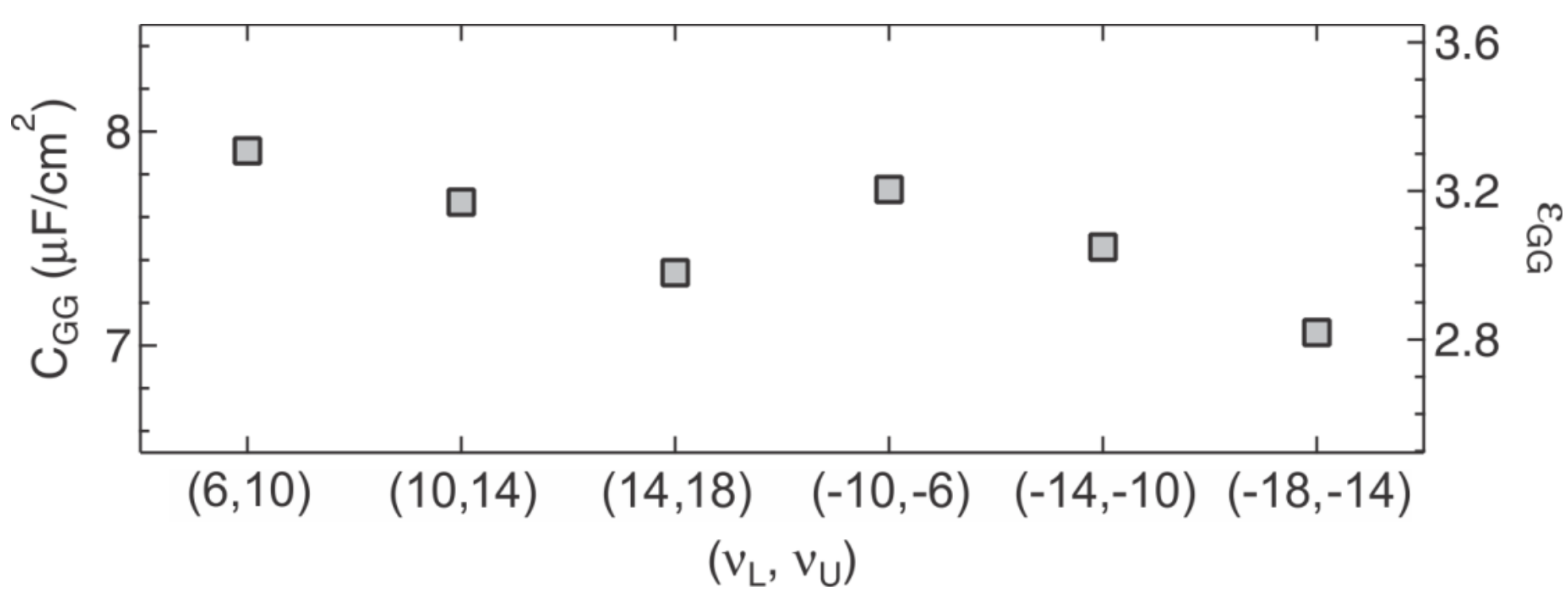

Figure S11. $C_{GG}$ and $\varepsilon_{GG}$ extracted from several LL crossings (with filling factors ($\nu_L, \nu_U$) from the lower and upper layers) in Device D (see Fig. S10). The $C_{GG}$ extracted from several LL crossings indicated in Fig. S10 are consistent with each other, with an average $C_{GG}$ = (7.5 ± 0.3) μF/cm$^2$, corresponding to $\varepsilon_{GG} \approx 2.9\,\varepsilon_0$ for $d_{GG}$ = 0.34 nm. The error in $C_{GG}$ denotes one standard deviation. Here to extract $C_{GG}$ we used the same procedure as in Device A (see Fig. S8(b) and also the main text for details).